\documentclass{siamltex}
\usepackage[numbers]{natbib}

\usepackage[a4paper]{geometry}
\usepackage{amsfonts,amsmath,amssymb}
\makeatletter
\newif\if@restonecol
\makeatother
\usepackage{natbib}
\usepackage{graphicx}
\usepackage{booktabs}
\usepackage{listings}
\usepackage{multirow}
\usepackage{amscd}
\usepackage{color}
\usepackage{url}
\usepackage[x11names, rgb]{xcolor}
\usepackage[utf8]{inputenc}
\usepackage{tikz}
\usetikzlibrary{matrix,arrows,decorations.pathmorphing,decorations.pathreplacing,positioning}
\usepackage{minibox}
\usepackage{textcomp}

\definecolor{gray}{rgb}{0.5,0.5,0.5}

\definecolor{dkgreen}{rgb}{0,0.6,0}
\definecolor{gray}{rgb}{0.5,0.5,0.5}
\definecolor{mauve}{rgb}{0.58,0,0.82}
\definecolor{lightyellow}{rgb}{255,254,230}

\title{A framework for the automation of generalised stability theory}

\author{P. E.\ Farrell\thanks{Applied Modelling and Computation Group, Department of Earth Science and Engineering, Imperial College London, London, UK, and Center for Biomedical Computing, Simula Research Laboratory, Oslo, Norway ({\tt patrick.farrell@imperial.ac.uk}).}
        \and C. J.\ Cotter\thanks{Department of Aeronautics, Imperial College London, London, UK ({\tt colin.cotter@imperial.ac.uk})}
        \and S. W.\ Funke\thanks{Applied Modelling and Computation Group, Department of Earth Science and Engineering, and Grantham Institute for Climate Change, Imperial College London, London, UK ({\tt s.funke09@imperial.ac.uk})}}

\begin{document}

  \maketitle

  \begin{abstract}
The traditional approach to investigating the stability of a physical system is to linearise the
equations about a steady base solution, and to examine the eigenvalues of the linearised operator.  Over
the past several decades, it has been recognised that this approach only determines the asymptotic
stability of the system, and neglects the possibility of transient perturbation growth arising due
to the nonnormality of the system. This observation motivated the development of a more powerful
generalised stability theory (GST), which focusses instead on the singular value decomposition of
the linearised propagator of the system. While GST has had significant successes in understanding
the stability of phenomena in geophysical fluid dynamics, its more widespread applicability has been
hampered by the fact that computing the SVD requires both the tangent linear operator and its
adjoint: deriving the tangent linear and adjoint models is usually a considerable challenge, and
manually embedding them inside an eigensolver is laborious. In this paper, we present a framework
for the automation of generalised stability theory, which overcomes these difficulties. Given a
compact high-level symbolic representation of a finite element discretisation implemented in the FEniCS system, 
efficient C++ code is automatically generated to assemble the forward, tangent linear and
adjoint models; these models are then used to calculate the optimally growing perturbations to the
forward model, and their growth rates. By automating the stability computations, we hope to make
these powerful tools a more routine part of computational analysis. The efficiency and generality of the
framework is demonstrated with applications drawn from geophysical fluid dynamics, phase separation
and quantum mechanics.
  \end{abstract}



  \pagestyle{myheadings}
  \thispagestyle{plain}
  \markboth{P. E.\ FARRELL, C. J.\ COTTER AND S. W.\ FUNKE}{THE AUTOMATION OF GENERALISED STABILITY THEORY}

  \setcounter{section}{0}
  \setcounter{equation}{0}

\section{Introduction}

The stability of a physical system is a classical problem
of mechanics, with contributions from authors such as Lagrange,
Dirichlet and Lyapunov \cite{leine2010}. Stability investigates 
the response of the system to small perturbations
applied to a particular initial condition: if for every $\epsilon$ there exists a $\delta$-neighbourhood of initial conditions
such that their solutions remain within the $\epsilon$-neighbourhood, then the system is stable at that initial condition; otherwise,
the system is unstable.

The traditional approach for investigating the stability of physical
systems was given by Lyapunov \cite{lyapunov1892}. The nonlinear
equations of motion are linearised about a base solution, and the eigenvalues
of the linearised system are computed. If all eigenvalues have negative real
part, then there exists a finite region of stability around the initial condition:
perturbations within that region decay to zero, and the system is asymptotically
stable \cite{parks1992}.

While this approach has had many successes, several authors have noted that
it does not give a complete description of the finite-time stability of a physical
system. While the eigendecomposition determines the asymptotic stability of the linearised equations as
$t \rightarrow \infty$, some systems permit transient perturbations which
grow in magnitude, before being predicted to decay. However, if the perturbations
grow too large, the linearised approximation may no longer be valid, and the system may become
unstable due to nonlinear effects. More specifically, this transient growth occurs when the system is nonnormal,
i.e. when the eigenfunctions of the system do not form an orthogonal basis \cite{schmid2007}.
For example, Trefethen et al. \cite{trefethen1993} describe how
the traditional approach fails to give accurate stability predictions
for several classical problems in fluid mechanics, and resolve the problem
by analysing the nonnormality of the system in terms of pseudospectra \cite{trefethen2006}.

Therefore, this motivates the development of a finite-time theory of stability,
to investigate and predict the transient growth of perturbations. While Lorenz \cite{lorenz1965}
discussed the core ideas (without using modern nomenclature), the development of this
so-called generalised stability theory (GST) has been driven by the work of
B. F. Farrell and co-workers (e.g., \cite{farrell1982a,farrell1985,farrell1996,farrell1996b}). The main idea
is to consider the linearised \emph{propagator} of the system, which is the operator (linearised
about the time-dependent trajectory) that maps perturbations in the initial conditions
to perturbations in the final state. Essentially, the propagator is the inverse of the tangent linear system associated
with the nonlinear forward model, along with operators to set the initial perturbation and select the final perturbation. The perturbations that grow maximally over the
time window are given by the singular functions of the propagator associated with
the largest singular values. Since the linearised propagator depends on the
base solution, it follows that the predictability of the system depends on the conditions
of the base solution itself: some states are inherently more predictable than others \cite{lorenz1965,kalnay2002}.
This idea has made a significant impact in the meteorological and oceanographic communities, 
and has been used to investigate many aspects of geophysical fluid dynamics \cite{lorenz1965,farrell1982a,farrell1985,palmer1988,moore2004,zanna2011,zanna2011b}.
In the fluid dynamics community, this technique is occasionally referred to
as direct optimal growth analysis \cite{barkley2008}.

While there are some applications of GST in other fields (e.g., \cite{davis2005,mao2011}), a large
number of the applications of this powerful idea have been in the area of geophysical fluid
dynamics. One reason for this is that the technique was invented in the meteorological community.
Another reason is that nonnormality is important in such flows, whereas traditional eigenvalue
analysis is sufficient for the normal case.  A final reason is that the necessary adjoint and
tangent linear models are commonly available in geophysical fluid dynamics, as they are necessary
components for variational data assimilation, whereas the difficulty of implementing them inhibits
the rapid application of GST in other scientific areas. Naumann \cite{naumann2011} describes the automatic
derivation of efficient adjoint and tangent linear models as ``one of the great open challenges of
High-Performance Scientific Computing''.

The main contribution of this work is a system for automating the calculations required to perform
a generalised stability analysis. Given a high-level description of a finite element discretisation
of the original time-dependent nonlinear model in the FEniCS framework \cite{logg2011}, a representation of the tangent linear and adjoint
models in the same high-level format are automatically derived at runtime \cite{farrell2012b}. These representations
are then passed to a finite element form compiler \cite{kirby2006}, which emits efficient C++ code for the assembly of
the nonlinear forward model, the tangent linear model, and its adjoint \cite{logg2010a}. The tangent linear and adjoint models are then used
automatically in a robust implementation of the Krylov-Schur algorithm \cite{hernandez2005} for computing a partial singular value decomposition of the
model propagator. By automating the difficult steps of deriving the tangent linear model and its adjoint, GST becomes much more
accessible: the analyst need only compactly describe a finite element discretisation of the problem of interest, and then can simply request
the fastest-growing perturbations and growth rates. The framework presented here is freely available under an open-source
license as part of the dolfin-adjoint package (\url{http://dolfin-adjoint.org}).

This paper is organised as follows. Section \ref{sec:gst} gives a brief overview of generalised stability theory, and
mentions some applications in the literature. Section \ref{sec:automation} describes the main contribution of this paper:
how the calculations involved in GST can be entirely automated. This relies on the automatic derivation of tangent linear
and adjoint models, as described in section \ref{sec:adjoints}. Finally, several examples are presented in section
\ref{sec:examples}. The examples are drawn from several areas of computational science to emphasise the widespread
applicability of the framework.

\section{Generalised stability theory} \label{sec:gst}
\subsection{The SVD of the propagator}
This presentation of generalised stability theory will consider the stability of the system to perturbations in
the initial conditions, but the same approach can be applied to analysing the stability of the system to
perturbations in other parameters.

Let $T$ be the time horizon of interest. Consider the solution of the model at the time $u_T$ as a pure function of the initial condition $u_0$:
\begin{equation} \label{eq:nonlinear_propagator}
u_T = M(u_0),
\end{equation}
where $M$ is the \emph{nonlinear propagator} that advances the solution in time over the given finite time window $[0, T]$.
Other parameters necessary for the solution (e.g. boundary conditions, material parameters, etc.)
are considered fixed. Assuming the model is sufficiently differentiable, the response of the 
model $M$ to a perturbation $\delta u_0$ in $u_0$ is given by
\begin{equation}
\delta u_T = M(u_0 + \delta u_0) - M(u_0) = \frac{\textrm{d} M}{\textrm{d} u_0} \delta u_0 +
\mathcal{O}\left(\left|\left|\delta u_0\right|\right|^2\right).
\end{equation}
Neglecting higher-order terms, the linearised perturbation to the final state is given by
\begin{equation}
\delta u_T \approx \frac{\textrm{d} M}{\textrm{d} u_0} \delta u_0 \equiv L \delta u_0,
\end{equation}
where $L$ is the \emph{linearised propagator} (or just propagator) ${\textrm{d} M}/{\textrm{d} u_0}$
that advances perturbations in the initial conditions to perturbations to the final solution. For example, if $u_T$ is the solution
of the linear ODE
\begin{equation}
  \frac{\mathrm{d} u}{\mathrm{d} t} = Au,
\end{equation}
then the propagator $L$ is given by
\begin{equation}
  L = e^{TA},
\end{equation}
where $e$ refers to the matrix exponential \cite{moler2003}. As discussed in \cite[figure 14.1]{trefethen2006}, 
the initial behaviour of $\left|\left|e^{TA}\right|\right|$ is governed by the numerical abscissa of $A$, while the asymptotic behaviour is governed by the spectrum of $A$. 
Techniques such as generalised stability analysis and pseudospectral analysis are most useful for intermediate values of $T$, which are of interest for transient growth.

To quantify the stability of the system, we wish to identify perturbations $\delta u_0$ that grow the most over the time window $[0, T]$.
For simplicity, equip both the initial condition and final solutions with the conventional inner product $\left\langle \cdot, \cdot \right\rangle$.
We seek the initial perturbation $\delta u^{\star}_0$ of unit norm $\left|\left|\delta u^{\star}_0\right|\right| = \sqrt{\left\langle \delta u^{\star}_0, \delta u^{\star}_0 \right\rangle} = 1$
such that
\begin{equation}
  \delta u^{\star}_0 = \operatorname*{arg\,max}_{\left|\left|\delta u_0\right|\right| = 1} \left\langle \delta u_T, \delta u_T \right\rangle.
\end{equation}
Expanding $\delta u_T$ in terms of the propagator,
\begin{equation}
\left\langle \delta u_T, \delta u_T \right\rangle = \left\langle L \delta u_0, L \delta u_0 \right\rangle = \left\langle \delta u_0, L^*L \delta u_0 \right\rangle,
\end{equation}
we see that the leading perturbation is the eigenfunction of $L^*L$ associated with the largest eigenvalue $\mu$, and the growth of the norm of the perturbation is given
by $\sqrt{\mu}$. In other words, the leading initial perturbation $\delta u^{\star}_0$ is the leading right singular function of $L$, the resulting final perturbation $\delta u_T$
is the associated left singular function, and the growth rate of the perturbation is given by the associated singular value $\sigma$. The remaining singular functions
offer a similar physical interpretation: if a singular function $v$ has an associated singular value $\sigma > 1$, the perturbation will grow over the finite time window $[0, T]$; if $\sigma < 1$,
the perturbation will decay over that time window. Note that the choice of $T$ is crucial: if $T$ is too large, the GST may predict contraction, even though significant transient growth may exist on
a shorter timescale.

If the initial condition and final solution spaces are equipped with inner products $\left\langle \cdot, \cdot \right\rangle_I \equiv \left\langle \cdot, X_I \cdot \right\rangle$ and
$\left\langle \cdot, \cdot \right\rangle_F \equiv \left\langle \cdot, X_F \cdot \right\rangle$ respectively, then the leading perturbations are given by the eigenfunctions
\begin{equation} \label{eqn:generalised_svd}
X_I^{-1} L^* X_F L \delta u_0 = \mu \delta u_0.
\end{equation}
The operators $X_I$ and $X_F$ must be symmetric positive-definite in order to define an inner
product. In the finite element context, $X_I$ and $X_F$ are often the mass matrices associated with
the input and output spaces, as these matrices induce the $L^2$ norm. All subsequent uses
of the term SVD in this paper are taken to include this generalised SVD \eqref{eqn:generalised_svd}.

\subsection{Computing the propagator}
In general, the nonlinear propagator $M$ that maps initial conditions to final solutions is not available as an explicit function;
instead, a PDE is solved. For clarity, let $m$ denote the data supplied for the initial condition. The PDE may be written in the abstract implicit
form
\begin{equation} \label{eq:pde}
F(u, m) = 0,
\end{equation}
with the understanding that $u_0 = m$. We assume that for any initial condition $m$, the PDE \eqref{eq:pde} can be solved for the solution trajectory $u$;
the nonlinear propagator $M$ can then be computed by returning the solution at the final time. Differentiating \eqref{eq:pde} with respect to the
initial condition data $m$ yields
\begin{equation} \label{eq:tlm}
  \frac{\partial F}{\partial u} \frac{\textrm{d}u}{\textrm{d}m} = - \frac{\partial F}{\partial m},
\end{equation}
the \emph{tangent linear system} associated with the PDE
\eqref{eq:pde}.  The term ${\partial F}/{\partial u}$ is the PDE
operator linearised about the solution trajectory $u$: therefore, it
is linear, even when the original PDE is nonlinear. ${\partial
  F}/{\partial m}$ describes how the equations change as the initial
condition data $m$ changes, and acts as the source term for the
tangent linear system. ${\textrm{d}u}/{\textrm{d}m}$ is the prognostic
variable of the tangent linear system \eqref{eq:tlm}, and describes
how the solution changes with changes to $m$. To evaluate the action
of the propagator $L$ on a given perturbation $\delta m$, the tangent
linear system is solved with that particular perturbation, and
evaluated at the final time:
\begin{equation}
L \delta m \equiv - \left.\left(\frac{\partial F}{\partial u}\right)^{-1}\frac{\partial F}{\partial m} \delta m\right|_T.
\end{equation}

Therefore, to automate the generalised stability analysis of a PDE \eqref{eq:pde}, it is necessary to automatically derive and solve
the associated tangent linear system \eqref{eq:tlm}. Furthermore, as discussed in section \ref{sec:svd}, all algorithms for computing
the SVD of a matrix $A$ require its adjoint $A^*$; therefore, it is also necessary to automatically derive and solve the adjoint
of the tangent linear system. If the PDE is linear and steady, then this derivation is straightforward; however, if the PDE
is nonlinear and time-dependent, the derivation of the associated tangent linear and adjoint systems is widely regarded as a major
challenge, even with the assistance of algorithmic differentiation tools \cite{naumann2011}. Another crucial concern is the efficiency
of the derived models: the SVD computation requires many runs of the tangent linear and adjoint systems, and so their computational
performance is of great importance if the stability analysis is to be tractable. However, by exploiting the special structure of
finite element discretisations, it is possible to entirely automate the derivation of efficient tangent linear and adjoint models; this
is the subject of the next section.

\section{Automating generalised stability theory} \label{sec:automation}
The following sections explain in detail how the SVD computation is automated by combining the FEniCS framework \cite{logg2011}, dolfin-adjoint and SLEPc \cite{hernandez2005,hernandez2007b}.

\begin{figure}
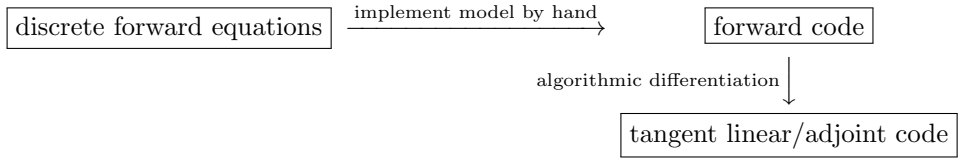

\begin{equation*}
  \begin{CD}
    \fbox{\textrm{discrete forward equations}}
    @>\textrm{implement model by hand}>>
    \fbox{\textrm{forward code}}\\
    @.
    @V\textrm{algorithmic differentiation}VV \\
    @.
    \fbox{\textrm{tangent linear/adjoint code}}\\
  \end{CD}
\end{equation*}
  \caption{The traditional approach to developing tangent linear and adjoint models. The forward model
  is implemented by hand, and its adjoint derived either by hand or (more often) with the assistance
  of an algorithmic differentiation tool.}
  \label{fig:traditional_approach}
\end{figure}

\subsection{Automating the generation of tangent linear and adjoint models} \label{sec:adjoints}
This section summarises the novel approach taken for deriving the tangent linear and adjoint models associated with a given
PDE solver. The main advantages over traditional approaches are its complete automation, its high performance, and its
trivial parallelisation. The approach is more fully described in \cite{farrell2012b}.

The traditional approach to automatically deriving the tangent linear and adjoint models associated with a given
PDE solver is to use algorithmic differentiation (AD, also known as automatic differentiation) tools  \cite{griewank2008,naumann2011}.
They primarily operate at the level of the source code (e.g., C++ or Fortran) that implements the discretisation, having already developed the
source code by hand. The main idea
is to treat the model as a (very long) sequence of elementary instructions, such as additions and multiplications, each of
which may be differentiated individually: the derived models are then composed using the chain rule applied forwards (in the
tangent linear case) or backwards (in the adjoint case). This approach is sketched in figure \ref{fig:traditional_approach}.

Naumann \cite{naumann2011} states that ``except for relatively simple cases, the differentiation
of computer programs is not automatic despite the existence of many reasonably mature AD software packages''.
This approach treats the model at a very low level of abstraction, and many of the difficulties of AD
stem from this fact. 

A source-to-source AD tool operating on the low-level code must parse the source to build a
representation of the sequence of elementary instructions as data.
This process is inherently fragile. The AD tool must handle complications such as preprocessor directives,
parallel directives, libraries for which the source code is not immediately available, expressions with
side effects, memory allocation, and aliasing. Correctly and efficiently handling these complications in
generality is very difficult, which puts a significant burden on both tool developers and users of algorithmic
differentiation.

However, with finite elements, it is possible to circumvent the problem of parsing source code. Finite element methods
are based on a powerful high-level abstraction: the language of variational forms. 
This mathematical abstraction naturally \emph{allows for the discrete
equations to be represented as data}. In the FEniCS project
\cite{logg2011}, the discrete variational form is represented in the Unified Form
Language (UFL) format, which is very similar to mathematical notation \cite{alnaes2011,alnaes2012}. This representation is then passed to a specialised finite element
form compiler \cite{kirby2006}, which emits optimised C++ code to assemble the desired discrete equations.
This approach has many advantages: it relieves the model developer of much of the manual labour (even complex
models such as the Navier-Stokes can be written in tens of lines of code), the form compiler can employ
specific optimisations that are complex to perform by hand \cite{olgaard2010}, and the generated code
can be tailored to the architecture at a very high level \cite{markall2012}.

In the context of stability analysis, this approach has one other major advantage: by representing the equations
to be solved as high-level data, the automated derivation of related models (such as the tangent linear and adjoint
systems) becomes much more tractable. This high-level abstraction for the finite element model matches naturally with a higher-level abstraction
for model differentiation: our approach takes the view that a model is a sequence of equation solves.
This approach is implemented in the dolfin-adjoint software package \cite{farrell2012b}. Its strategy for deriving the tangent
linear and adjoint models is now discussed.

When the dolfin-adjoint module is imported, all functions that solve equations or modify variable values are overloaded. 
In addition to providing their regular functionality, these overloaded functions build
a \emph{tape} of the forward model at runtime: the tape records all details of the forward evaluation necessary for
evaluating the model at a different parameter. In low-level algorithmic differentiation, this consists of
a record of all elementary operations performed, along with their arguments \cite{griewank2008}; analogously, in the dolfin-adjoint case,
the tape records all forward equations solved (in UFL format), their boundary conditions, their dependencies on previously computed values, etc.
The tape contains a complete record of the discrete forward model, and may be used to re-execute the forward model, which
finds applications in PDE-constrained optimisation and checkpointing.

The tape contains all information necessary to derive the tangent linear and adjoint models associated with the discrete forward model.
For concreteness, consider the derivation of the
tangent linear model. Each equation in the forward model induces an associated equation in the tangent linear model.
Let $u_k$ be the variable solved for in equation $k$ of the forward model. Suppose the forward equation may be written
as
\begin{equation} \label{eqn:sample_fwd}
  F_k(u_k, u_{k_1}, \dots, u_{k_N}) = 0,
\end{equation}
where $F_k$ is a (possibly nonlinear) operator, and $u_{k_1}, \dots, u_{k_N}$ are $N$ previously computed values on which the equation depends ($k_i < k \ \forall i$).
Let $\delta{m}$ be a perturbation to $m$ whose impact is to be quantified.
By differentiating \eqref{eqn:sample_fwd} with respect to $m$, we obtain the associated tangent linear equation
\begin{equation} \label{eqn:sample_tlm}
  \frac{\partial F_k}{\partial u_k} \dot{u}_k = \sum_{i=1}^{N} -\frac{\partial F_k}{\partial u_{k_i}} \dot{u}_{k_i},
\end{equation}
where 
\begin{equation}
  \dot{u}_j \equiv \dfrac{\mathrm{d}u_j}{\mathrm{d}m} \delta{m}
\end{equation}
is the tangent linear solution associated with $u_j$. If any boundary conditions are imposed strongly
on \eqref{eqn:sample_fwd}, their homogenised counterparts are imposed strongly on \eqref{eqn:sample_tlm}; weakly imposed
boundary conditions are handled naturally in the formulation. Note that $\dot{u}_{k_1}, \dots, \dot{u}_{k_N}$ must be computed before 
the equation for $\dot{u}_k$ may be assembled, in the same way that $u_{k_1}, \dots, u_{k_N}$ must be computed before the equation for
$u_k$ may be assembled. Since (i) the tape represents $F_k$ symbolically in UFL format, (ii) the tape records which variables
$u_{k_1}, \dots, u_{k_N}$ equation $k$ depends on, and (iii) UFL supports the symbolic differentiation
of operators with respect to their dependencies, the tangent linear equation \eqref{eqn:sample_tlm} may be derived by symbolic manipulation of the
data stored on the tape for the forward equation \eqref{eqn:sample_fwd}. 
Although the adjoint case is more complex, the associated adjoint equation may similarly be
derived by symbolic manipulation of the tape; for full details, see \cite{farrell2012b}.

By coupling the high-level representation of the forward model with this high-level differentiation approach,
the tangent linear and adjoint versions of a model written in the FEniCS framework may be derived with almost no user
intervention or effort \cite{farrell2012b}: this is because all of the necessary manipulation steps are fully automatable when
the tape retains the symbolic structure of the equations. With the dolfin-adjoint software package, the discrete tangent linear and adjoint equations
to be solved are symbolically derived in the exact same UFL format as the forward model, and passed to the same finite element
compiler.

This alternative approach to automating the derivation of the tangent linear and adjoint models has several major
advantages for generalised stability analysis. Firstly, the derivation of the tangent linear and adjoint models is
almost entirely automatic. In the example shown in section \ref{sec:code_example}, the user need only add two lines of
code: one to import the dolfin-adjoint library, and one to request the leading singular triplets. Secondly, the derived
tangent linear and adjoint models approach optimal theoretical efficiency. This is crucial, as the
SVD calculation requires many iterations of the tangent linear and adjoint models; the efficiency of the approach will be
demonstrated on several examples in section \ref{sec:examples}. Thirdly, whereas applying algorithmic
differentiation to a parallel code is a major research challenge \cite{utke2009,forster2011}, this high-level approach
parallelises very naturally: if the forward model runs in parallel, the tangent linear and adjoint models will also \cite{farrell2012b}. In
fact, there is no parallel-specific code in dolfin-adjoint -- by operating on the discrete
equations instead of the source code, the problem of parallelisation dissolves. As the computational demands in problems
of practical interest are usually very large, parallelisation is a necessity if the GST framework is to be used in such
cases.

\subsection{Singular value decomposition} \label{sec:svd}
Once the propagator $L$ is available, its singular value decomposition may be computed. 
There are two main computational approaches. The first approach is to compute the eigendecomposition 
of the \emph{cross product} matrix $L^*L$ (or $LL^*$, whichever is smaller). The second is to
compute the eigendecomposition of the \emph{cyclic} matrix 
\begin{equation}
H(L) = 
\begin{pmatrix} 0 & L \\
              L^* & 0
\end{pmatrix}.
\end{equation}
The latter option is more accurate for computing the
small singular values, but is more expensive \cite{trefethen1997}. As we are only interested in a small number of
the largest singular triplets, the cross product approach is used throughout this work. Note
that regardless of which approach is taken, the adjoint propagator $L^*$ is necessary
to compute the SVD of $L$.

The algorithm used to compute the eigendecomposition of the cross product matrix is the Krylov-Schur algorithm
\cite{stewart2001}, as
implemented in SLEPc \cite{hernandez2005,hernandez2007b}. As the cross product matrix is Hermitian, this
algorithm reduces to the thick-restart variant \cite{wu2000} of the Lanczos method \cite{lanczos1950}.  This algorithm was found experimentally to be
faster than all other algorithms implemented in SLEPc for the computation of a small number of singular triplets, which
is the case of interest in stability analysis.

Rather than computing and storing a dense matrix representation of the propagator, the action of the propagator is computed in a matrix-free
fashion, using the tangent linear model. In turn, the entire time-dependent tangent linear model is not
stored, but its action is implemented as the solution of several equations in sequence.
In turn, the solution of each equation may optionally be achieved in a matrix-free fashion; the automatic
derivation of the tangent linear and adjoint systems supports such an approach \cite{farrell2012b}.
Similarly, the adjoint propagator is computed in a matrix-free fashion using the adjoint model. SLEPc
elegantly supports such matrix-free computations through the use of PETSc shell matrices \cite{balay2010,balay1997}.

\subsection{Code example and implementation} \label{sec:code_example}

\begin{figure}
\centering
{\normalsize
\begin{lstlisting}[language=Python,
numbers=left,numberstyle=\tiny\color{gray},
morekeywords={UnitInterval,FunctionSpace,project,Expression,TestFunction,Constant,grad,dx,DirichletBC,solve,assign,norm,plot,inner,show},
emph={assign,solve,Function},
emphstyle=\color{red}\textbf,
emph={[2]dolfin_adjoint,compute_gst},
emphstyle={[2]\color{blue}\textbf}]
from dolfin import *
# Import the dolfin-adjoint library to enable the derivation 
# of tangent linear and adjoint models  
from dolfin_adjoint import *

# Define the computational domain and function spaces
mesh = UnitInterval(10)
V = FunctionSpace(mesh, "Lagrange", 1)

# Define the necessary test and trial functions
ic = project(Expression("sin(2*pi*x[0])"),  V)
u = Function(ic, name="State")
u_next = Function(V, name="NextState")
v = TestFunction(V)

# Define the viscosity and timestep
nu = Constant(0.0001)
timestep = Constant(0.1)

# Define the weak formulation of the Burgers' equation
F = (((u_next - u)/timestep)*v
     + u_next*grad(u_next)*v + nu*grad(u_next)*grad(v))*dx
bc = DirichletBC(V, 0.0, "on_boundary")

# Run the forward model
t = 0.0
end = 0.2
while (t <= end):
    solve(F == 0, u_next, bc)
    u.assign(u_next)

    t += float(timestep)

# Compute the five largest singular values for the propagator
# that maps the initial state of the Burgers' solution 
gst = compute_gst(ic="State", final="State", nsv=5)
\end{lstlisting}
}
\caption{The entire code to compute a generalised stability analysis of the nonlinear Burgers'
  equation. This code uses piecewise linear Lagrange finite elements for the spatial discretisation
  (lines 8, 21--22), and implicit Euler for the temporal discretisation (lines 21--22). The
  high-level approach leads to extremely compact and readable code. In order to use the framework
  presented here, only two additional lines are necessary (in blue): one to import the
  dolfin-adjoint library (line 4), and one to compute the singular value decomposition of the propagator
  associated with the forward model (line 36). The functions in red are overloaded by dolfin-adjoint in order
  to record the information necessary for the derivation of the tangent linear and adjoint models as
  described in section \ref{sec:adjoints}. The \texttt{compute\_gst} function (line 36) symbolically derives
  the tangent linear and adjoint models, creates a shell matrix to compute the action of the
propagator, and embeds it inside a Krylov-Schur algorithm to compute the requested number of
singular triplets.}
\label{fig:code_example}
\end{figure}

\begin{figure}
\centering
\begin{tikzpicture}[>=angle 90]
\matrix(a)[matrix of math nodes, row sep=3em]
{
 \framebox{\minibox{\hfill forward model \hspace*{\fill}\\ \hfill (symbolic representation) \hspace*{\fill}}}
&&
 \fbox{\minibox{\hfill forward model \hspace*{\fill}\\\hfill (code)\hspace*{\fill}}}
\\
\framebox{\minibox{\hfill adjoint model \hspace*{\fill}\\ \hfill (symbolic representation) \hspace*{\fill}}}
&&
\fbox{\minibox{\hfill tangent linear model \hspace*{\fill}\\\hfill (symbolic representation)\hspace*{\fill}}}
\\
\framebox{\minibox{adjoint model \\ \hfill (code) \hspace*{\fill}}}
&&
\fbox{\minibox{tangent linear model\\\hfill (code)\hspace*{\fill}}}
\\
& \fbox{\minibox{SVD}}
&\\
};
\path[->](a-1-1) edge node[left]{dolfin-adjoint} (a-2-1);
\path[->](a-1-1) edge node[above]{FEniCS} (a-1-3);
\path[->](a-1-1) edge node[above right]{dolfin-adjoint} (a-2-3);
\path[->](a-2-1) edge node[left]{FEniCS} (a-3-1);
\path[->](a-2-3) edge node[right]{FEniCS} (a-3-3);
\path[->](a-3-1) edge node[below left]{SLEPc} (a-4-2);
\path[->](a-3-3) edge node[below right]{SLEPc} (a-4-2);
\end{tikzpicture}
\caption{
The software components for computing the SVD. 
    The user specifies the discrete forward equations in a
    high-level language similar to mathematical notation; the discrete
    forward equations are explicitly represented in memory in the UFL
    format. The in-memory representation of the associated
    tangent linear and adjoint systems is derived by dolfin-adjoint 
    from the in-memory representation of the forward
    problem.  Both the forward and adjoint equations are then passed
    to the FEniCS system, which automatically generates and executes
    the code necessary to compute the forward and adjoint solutions.
Finally, SLEPc is used to compute the singular value decomposition. 
}\label{fig:software_components}
\end{figure}
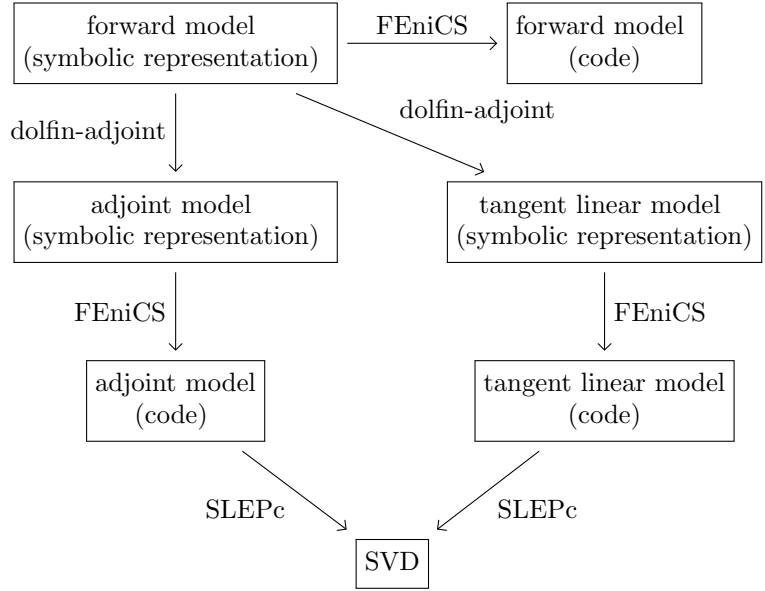

In order to demonstrate the user interface of the proposed framework, a code example for a generalised stability analysis of the nonlinear Burgers'
equation is given in figure \ref{fig:code_example}. The example is complete; nothing has been removed. Only two lines
of code are added to the forward model to conduct the GST: one to import the dolfin-adjoint library, and one to perform
the GST computation. 

We now discuss the internals of the \texttt{compute\_gst} function (figure \ref{fig:software_components}). The computation of the eigendecomposition
is driven by SLEPc via the \texttt{EPSSolve} function. The main input to this routine is a PETSc shell matrix that represents
the GST operator
\begin{equation}
G = X_I^{-1} L^* X_F L.
\end{equation}
If no $X_I$ or $X_F$ are specified, the mass matrices of the associated function spaces are used by default. This shell matrix
is equipped with a function that computes its action, by composing the action of the four constituent matrices.
The computation of the actions of $X_F$ and $X_I^{-1}$ are straightforward and are not discussed further.

The propagator $L$ is in turn represented as a shell matrix equipped with two operations, one for its
action and one for its Hermitian action.  The action of $L$ on a vector $\delta u$ is computed by inspecting
the tape built by dolfin-adjoint during the initial forward run and deriving the tangent linear equation associated with each equation of the
forward model, as discussed in section \ref{sec:adjoints}. The tape represents each forward equation symbolically in UFL format; the derived tangent
linear equations are represented in the same UFL format, which means that efficient code for their assembly
can be generated using the FEniCS system via automated code generation and just-in-time compilation. Each equation of the tangent linear system is
solved in turn, with the source term $\delta u$ added to the right-hand-side of the tangent linear equation
associated with the forward variable that is defined to be the input of the propagator. Once the tangent
linear equation associated with the output forward variable is solved, the tangent linear solution is returned
as the action of the propagator.

The same strategy is used to compute the Hermitian action of the propagator, \emph{mutatis mutandis}.
The adjoint equations are solved in the opposite order to that of the forward model. Again, the assembly of the
adjoint equations relies on the automated code generation technology of FEniCS. The perturbation on
which $L^*$ is acting is added to the right-hand-side of the adjoint equation associated with the output variable of the propagator, and the adjoint solution
associated with the input variable of the propagator is returned.

This implementation strategy has several advantages. SLEPc cleanly separates the algorithm for computing the
eigendecomposition from the implementation of the matrices representing the propagator $L$ and the GST
operator $G$.  This means that developments in SLEPc (such as new algorithms, or improvements to existing ones) are
immediately available.  By exploiting the code generation facilities of FEniCS to implement the tangent linear
and adjoint models, the implementation inherits all of its advantages, such as parallelism,
efficiency and generality.  Finally, by relying on dolfin-adjoint for the automated derivation of the tangent
linear and adjoint models, the user is relieved of the burden of manually deriving, implementing and
maintaining them. 

In combination, this system allows for the flexible and efficient computation of generalised
stability analyses, so long as the forward model is representable in the FEniCS system. The FEniCS system
supports a wide variety of finite elements (including arbitrary order continuous and discontinuous Lagrange, Raviart--Thomas, Nédelec,
Brezzi--Douglas--Marini, Crouzeix--Raviart \cite{kirby2004}), distributed-memory unstructured meshes of complicated geometries,
and any finite element discretisation that can be represented in UFL. This includes sophisticated
discretisations of complex PDEs, including Stokes with nonlinear rheology for mantle convection
\cite{vynnytska2013}, viscoelastic deformation \cite{rognes2010b}, the
Landau--Lifshitz--Gilbert equation for micromagnetic simulations \cite{abert2013}, and the coupled
PDEs-ODEs describing the calcium release unit of sarcoplasmic reticulum in the heart \cite{hake2013}.

\section{Verification and applications} \label{sec:examples}
All applications are available under an open-source license as part of the dolfin-adjoint applications
repository (\url{http://dolfin-adjoint.org}). In all of the examples, the mass matrices of the input
and output spaces were used to define the norms in equation \eqref{eqn:generalised_svd}.
The benchmark tables show the minimum time of five experiments, performed on 8 2.13 GHz Intel Xeon CPU cores with 12 GB memory.

\subsection{Verification: the nonlinear Burgers' equation}
The verification of the framework proceeds in two stages. Firstly, the correctness of the tangent linear
and adjoint models must be verified. Secondly, the correctness of the singular value decomposition
must be verified.

The fundamental tool in verifying the correctness of the tangent linear and adjoint
models is the Taylor remainder test. Suppose we have a black box for evaluating a function $f(x)$,
and have a candidate function for its gradient $\nabla f$. The correctness of the gradient can
be asserted by noting that by Taylor's theorem, the first order Taylor remainder
\begin{equation}
  \left| f(x + h \delta x) - f(x) \right| \rightarrow 0 \quad \textrm{at} \  O(h)
\end{equation}
converges to zero at first order, but that the Taylor remainder corrected with the gradient
\begin{equation} \label{eq:taylor_2nd}
  \left| f(x + h \delta x) - f(x) - h\delta x^T \nabla f \right| \rightarrow 0 \quad \textrm{at} \  O(h^2)
\end{equation}
converges to zero at second order. In this context, the function $f(u)$ is a functional of the solution $u$
of a PDE system $F(u,m) = 0$ specified by parameters $m$, and its gradient $\nabla_m f(u(m))$ is computed in two different ways, once using the tangent linear model
and once using its adjoint.

For the verification exercise, we choose as our model the nonlinear time-dependent Burgers'
equation:
\begin{equation} \label{eq:burgers}
  F(u, m) \equiv \frac{\partial u}{\partial t} + u \cdot \nabla u - \nu \nabla^2 u = 0,
\end{equation}
on some domain $\Omega \times [0, T]$, along with suitable boundary conditions and diffusivity coefficient $\nu$. The parameter $m$ is the initial condition
for $u$. We choose our functional $J$ as
\begin{equation}
  J(u) = \int_{\Omega} \left| u_T \right|^2 \ \textrm{d}x,
\end{equation}
the square of the $L^2$ norm of the solution evaluated at the end of time. By the chain rule, the gradient
${\textrm{d}J(u(m))}/{\textrm{d}m}$ can be computed with
\begin{equation}
  \frac{\textrm{d}J(u(m))}{\textrm{d}m} = \left \langle \frac{\partial J}{\partial u}, \frac{\textrm{d}u}{\textrm{d}m} \right \rangle,
\end{equation}
where ${\textrm{d}u}/{\textrm{d}m}$ is the solution of the associated tangent linear system \eqref{eq:tlm}.
In this way, the automated derivation of the tangent linear system \eqref{eq:tlm} from the nonlinear
forward model \eqref{eq:burgers} can be rigorously verified: the tangent linear solution is correct
if and only if the second order Taylor remainder \eqref{eq:taylor_2nd} converges at second order.
In practice, computing the whole of the solution Jacobian ${\textrm{d}u}/{\textrm{d}m}$ is unnecessary, as we only require
the \emph{action} of the gradient ${\textrm{d}J}/{\textrm{d}m}$ on a particular perturbation $h \delta m$. In this case,
it is sufficient to compute
\begin{equation}
  \left\langle \frac{\textrm{d}J(u(m))}{\textrm{d}m}, h \delta m \right \rangle = \left \langle \frac{\partial J}{\partial u}, h \frac{\textrm{d}u}{\textrm{d}m} \delta m\right \rangle,
\end{equation}
where the action of the solution Jacobian ${\textrm{d}u}/{\textrm{d}m}$ on the perturbation $h \delta m$ is computed via
\begin{equation}
  \frac{\partial F}{\partial u} \left(h \frac{\textrm{d}u}{\textrm{d}m} \delta m\right) = - h \frac{\partial F}{\partial m} \delta m.
\end{equation}

\begin{table}
\centering
\begin{tabular}{ccccc}
\toprule
$h$ & \small{$\left|\widehat{J}(\tilde{m}) - \widehat{J}(m_0) \right|$} & order & \small{$\left|\widehat{J}(\tilde{m}) - \widehat{J}(m_0) - \tilde{m}^T \nabla \widehat{J} \right|$} & order \\
\midrule
$1 \times 10^{-3}$ & 1.8664 $\times 10^{-5}$ &    & 5.8991 $\times 10^{-7}$ & \\
$5 \times 10^{-4}$ & 9.4796 $\times 10^{-6}$  & 0.9773 & 1.4747 $\times 10^{-7}$ & 2.000 \\
$2.5 \times 10^{-4}$ & 4.7766 $\times 10^{-6}$ & 0.9888 &  3.6868 $\times 10^{-8}$ & 2.000 \\
$1.25 \times 10^{-4}$ & 2.3975 $\times 10^{-6}$ & 0.9944 & $9.2169 \times 10^{-9}$ & 2.000 \\
$6.25 \times 10^{-5}$ & 1.2010 $\times 10^{-6}$ & 0.9972 &  $2.3042 \times 10^{-9}$ & 2.000 \\
\bottomrule
\end{tabular}
\caption{Verification of the tangent linear model. The Taylor remainders for the functional
  $\widehat{J} = J(u(m))$ are evaluated at a perturbed initial condition
  $\tilde{m} \equiv m_0 + h\delta m$, where the perturbation
  direction $\delta m$ is pseudorandomly generated. As expected, the
  Taylor remainder incorporating gradient information computed using
  the tangent linear model converges at second order, indicating that
  the functional gradient computed using the tangent linear model is
  correct.}
\label{tab:burgers_tlm}
\end{table}
The Burgers' equation \eqref{eq:burgers} is discretised in space using standard piecewise quadratic finite elements and discretised in time
using the trapezoidal rule, and the resulting nonlinear system solved via Newton iteration. As described in section \ref{sec:automation}, the tangent linear model
is automatically derived, with almost no user intervention. The results of the Taylor remainder test for the
tangent linear model can be seen in table \ref{tab:burgers_tlm}. As expected, the Taylor remainders corrected with
the functional gradient do indeed converge at second order, indicating that the computed gradient, the tangent linear
solution, and the tangent linear equations are all correct.

Similarly, the adjoint model may be verified, by computing the gradient ${\textrm{d}J}/{\textrm{d}m}$ via
the relation
\begin{equation}
  \frac{\textrm{d}J(u(m))}{\textrm{d}m} = -\left \langle \lambda, \frac{\partial F}{\partial m} \right \rangle,
\end{equation}
where $\lambda$ is the solution of the adjoint equation
\begin{equation} \label{eq:adj}
  \left(\frac{\partial F}{\partial u}\right)^* \lambda = \frac{\partial J}{\partial u}^*.
\end{equation}

\begin{table}
\centering
\begin{tabular}{ccccc}
\toprule
$h$ & \small{$\left|\widehat{J}(\tilde{m}) - \widehat{J}(m_0) \right|$} & order & \small{$\left|\widehat{J}(\tilde{m}) - \widehat{J}(m_0) - \tilde{m}^T \nabla \widehat{J} \right|$} & order \\
\midrule
$1 \times 10^{-3}$ & 4.0880 $\times 10^{-5}$ &    & 9.5164 $\times 10^{-7}$ & \\
$5 \times 10^{-4}$ & 2.0678 $\times 10^{-5}$  & 0.9833 & 2.3786 $\times 10^{-7}$ & 2.000 \\
$2.5 \times 10^{-4}$ & 1.0398 $\times 10^{-5}$ & 0.9917 &  5.9459 $\times 10^{-8}$ & 2.000 \\
$1.25 \times 10^{-4}$ & 5.2141 $\times 10^{-6}$ & 0.9958 & 1.4864 $\times 10^{-9}$ & 2.000 \\
$6.25 \times 10^{-5}$ & 2.6107 $\times 10^{-6}$ & 0.9979 &  3.7159 $\times 10^{-9}$ & 2.000 \\
\bottomrule
\end{tabular}
\caption{Verification of the adjoint model. The Taylor remainders for the functional
  $\widehat{J} = J(u(m))$ are evaluated at a perturbed initial condition
  $\tilde{m} \equiv m_0 + h\delta m$, where the perturbation
  direction $\delta m$ is pseudorandomly generated. As expected, the
  Taylor remainder incorporating gradient information computed using
  the adjoint model converges at second order, indicating that
  the functional gradient computed using the adjoint model is
  correct.}
\label{tab:burgers_adj}
\end{table}

The results of the Taylor remainder test for the
adjoint model can be seen in table \ref{tab:burgers_adj}. Again, the Taylor remainders corrected with
the functional gradient do indeed converge at second order, indicating that the computed gradient, the adjoint
solution, and the automatically derived adjoint equations are all correct.

With the correctness of the tangent linear and adjoint models established, the correctness of
the singular value decomposition was verified. As described in section \ref{sec:svd}, in practical
computations the propagator is never represented as a matrix: instead, its action is computed
using the tangent linear model. However, for verification purposes, a dense matrix representation $U\Sigma V^*$ was
computed by performing the full singular value decomposition
of the propagator and multiplying the output matrices together.
(This calculation was expensive, and unnecessary in the general
case: the computation was performed merely for the purposes of verification). The action of this dense matrix
was compared against the matrix-free action with the tangent linear model on hundreds of random vectors $t$ (with
each component drawn from $\mathcal{U}\left(0, 1\right)$), by asserting that
\begin{equation}
  \left|\left|(U \Sigma V^*) t - Lt\right|\right| < \epsilon
\end{equation}
for each $t$, with $\epsilon = 10^{-7}$.
Additionally, the matrix-free action of $L$ was computed on each right singular vector $v$, and the
result compared to the prediction of the associated left singular vector from the singular value decomposition, by asserting that
\begin{equation}
  \left|\left|u - Lv\right|\right| < \epsilon
\end{equation}
for each $v$.

Finally, the relevance of the computed SVD was verified by running the nonlinear forward model
with the initial condition perturbed with the leading right singular vector. The actual growth rate of the perturbation
was compared with the growth rate predicted from the singular value; the prediction matched the actual
growth rate to within 1\%. This confirms the physical utility of the SVD for predicting the dynamics of small perturbations
to the initial condition.

\subsection{Navier-Stokes: double-diffusive salt fingering}

\begin{figure}
  \centering
  \begin{tabular}{cc}
    \includegraphics[width=5.0cm]{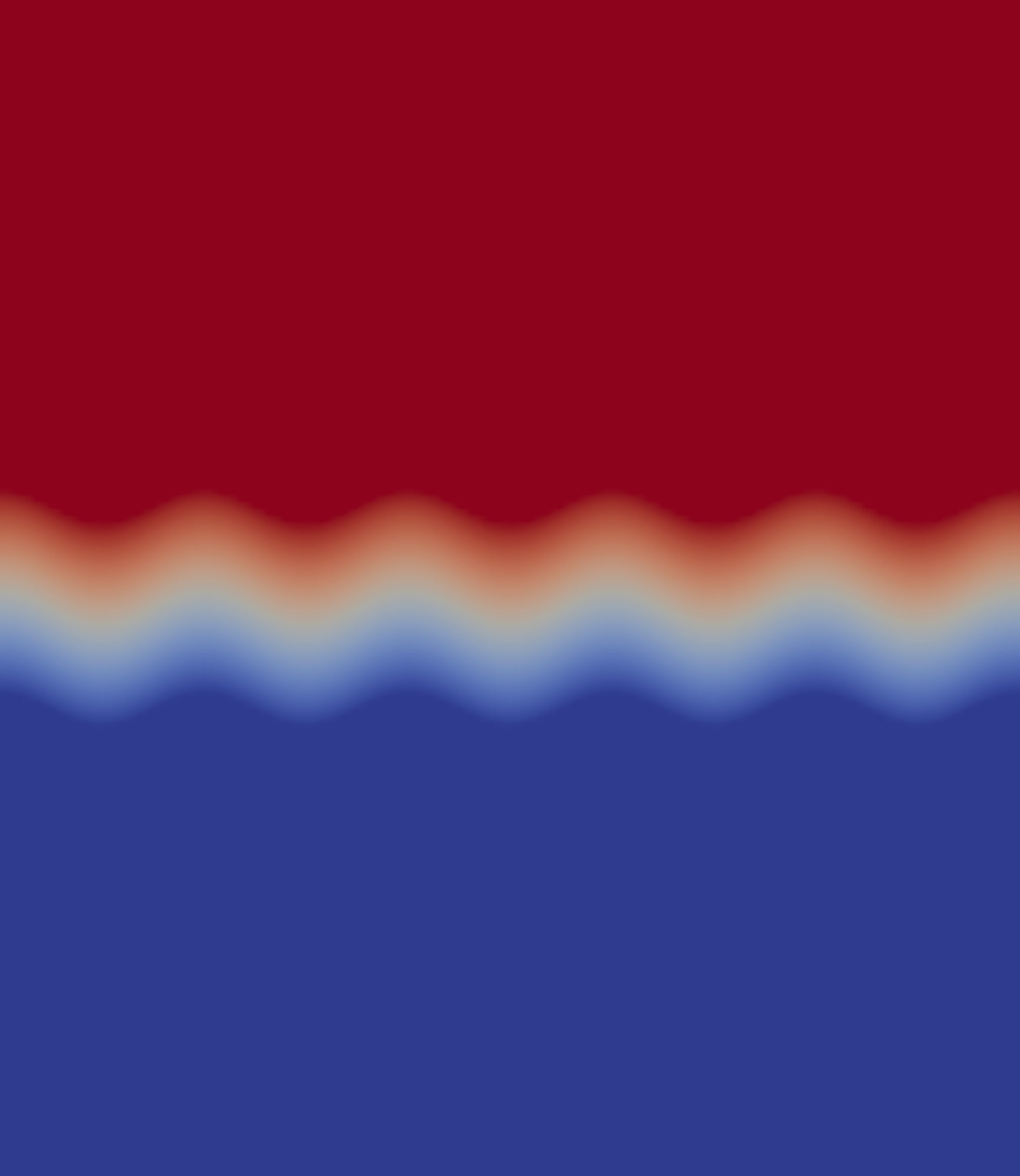} & \includegraphics[width=5.0cm]{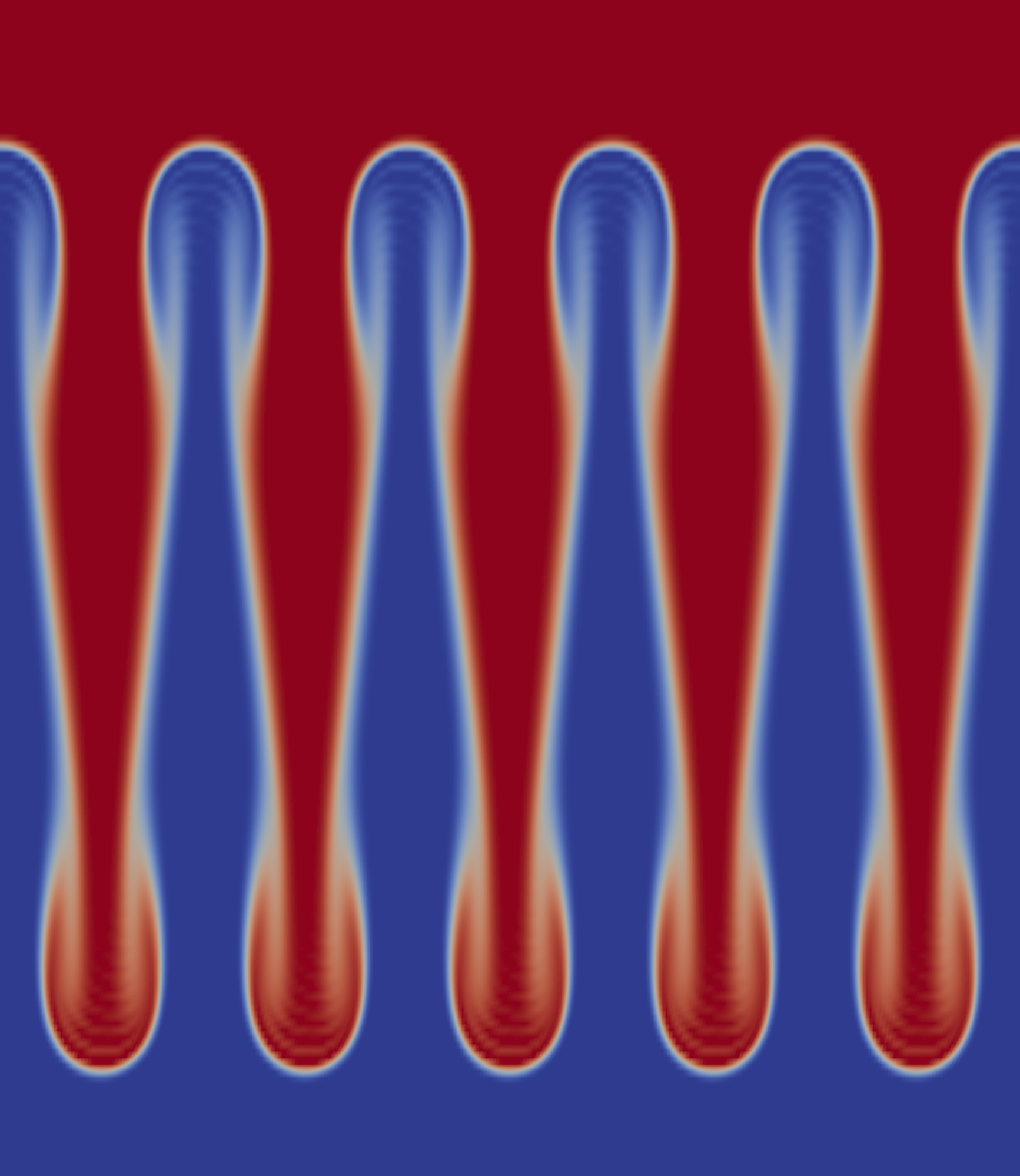}
    \\
    initial salinity & final salinity
  \end{tabular}
  \caption{The phenomenon of salt fingering. Warm salty water overlies cold fresh water. If a parcel of warm salty water
  sinks downwards into the colder region, the heat of the parcel is diffused away much faster than its salt, thus making
  the parcel denser, and causing it to sink further. Left: the initial condition for salinity, using the perturbed interface of \cite{ozgokmen1998b}. Right:
  the final salinity, at $T=0.05$.}
  \label{fig:salt_fingering}
\end{figure}

In the ocean, the diffusivity coefficient of temperature is approximately two orders of magnitude larger than
the diffusivity coefficient of salinity. Suppose warm salty water lies above colder, less salty water. If a parcel of warm salty water sinks downwards
into the colder region, the heat of the parcel will diffuse away much faster than its salt, thus making the parcel
denser, and causing it to sink further. Similarly, if a parcel of cold, less salty water rises into the warmer
region, it will gain heat from its surroundings much faster than it will gain salinity, making the parcel more
buoyant. This phenomenon is referred to as ``salt fingering'' \cite{stern1960} (figure \ref{fig:salt_fingering}) and has been observed in many
real-world oceanographic contexts \cite{turner1985}. An initial investigation of this phenomenon using the tools
of generalised stability theory was presented in \cite{eisenman2005}.

\newcommand{\Ra}{\textrm{Ra}}
\newcommand{\Sc}{\textrm{Sc}}
\renewcommand{\Pr}{\textrm{Pr}}

\"Ozg\"okmen and Esenkov \cite{ozgokmen1998b} used a numerical model to investigate asymmetry in the growth of salt fingers caused by
nonlinearities in the equation of state. In this work, we investigate the stability of the
proposed configuration to small perturbations, and examine what this means for its utility as a
numerical benchmark. The two-dimensional vorticity-streamfunction formulation of the
Navier-Stokes equations is coupled to two advection equations for temperature and salinity:
\begin{align}
  \frac{\partial \zeta}{\partial t} + \nabla^{\perp} \psi \cdot \nabla \zeta &= \frac{\Ra}{\Pr}\left(\frac{\partial
T}{\partial x} - \frac{1}{R_{\rho}^0} \frac{\partial S}{\partial x}\right) + \nabla^2 \zeta, \\
\frac{\partial T}{\partial t} + \nabla^{\perp} \psi \cdot \nabla T &= \frac{1}{\Pr} \nabla^2 T, \\
          \frac{\partial S}{\partial t} + \nabla^{\perp} \psi \cdot \nabla S &= \frac{1}{\Sc} \nabla^2 S, \\
                                       \nabla^2 \psi &= \zeta,
\end{align}
where $\zeta$ is the vorticity, $\psi$ is the streamfunction, $T$ is the temperature, $S$ is the
salinity, and $\Ra$, $\Sc$, $\Pr$ and ${R_{\rho}^0}$ are nondimensional parameters.
Periodic boundary conditions are applied on the left and right boundaries; for full details of the
remaining boundary conditions and values of the numerical parameters, see \cite{ozgokmen1998b}. The
configuration consists of two well-mixed layers (i.e., of homogeneous temperature and salinity) separated
by an interface. To activate the instability, \cite{ozgokmen1998b} added a
sinusoidal perturbation to the initial salinity field (figure \ref{fig:salt_fingering}). 

To investigate the possibility of a secondary instability about this perturbed initial condition,
the framework of generalised stability theory was applied. The PDE was discretised in space using
standard piecewise linear finite elements, and first-order $\theta$-timestepping was employed in time with
$\theta=0.6$. This value of $\theta$ was chosen to damp the over- and undershoots associated with the Galerkin advection of salinity \cite[\S 5]{davies2005}; an improved
implementation would use a more sophisticated advection scheme. At each timestep, the entire discretised nonlinear system was solved with
Newton iteration. The solution trajectory was computed using the initial sinusoidal perturbation to the
salinity field, the propagator was linearised about that trajectory, and the leading ten singular
triplets were computed. This calculation was repeated on several refinements of a structured mesh,
up to 300 $\times$ 300 cells, and with timesteps ranging from $1 \times 10^{-3}$ to $1.25 \times 10^{-4}$.

\begin{figure}
  \centering
  \begin{tabular}{cc}
    \includegraphics[width=5.0cm]{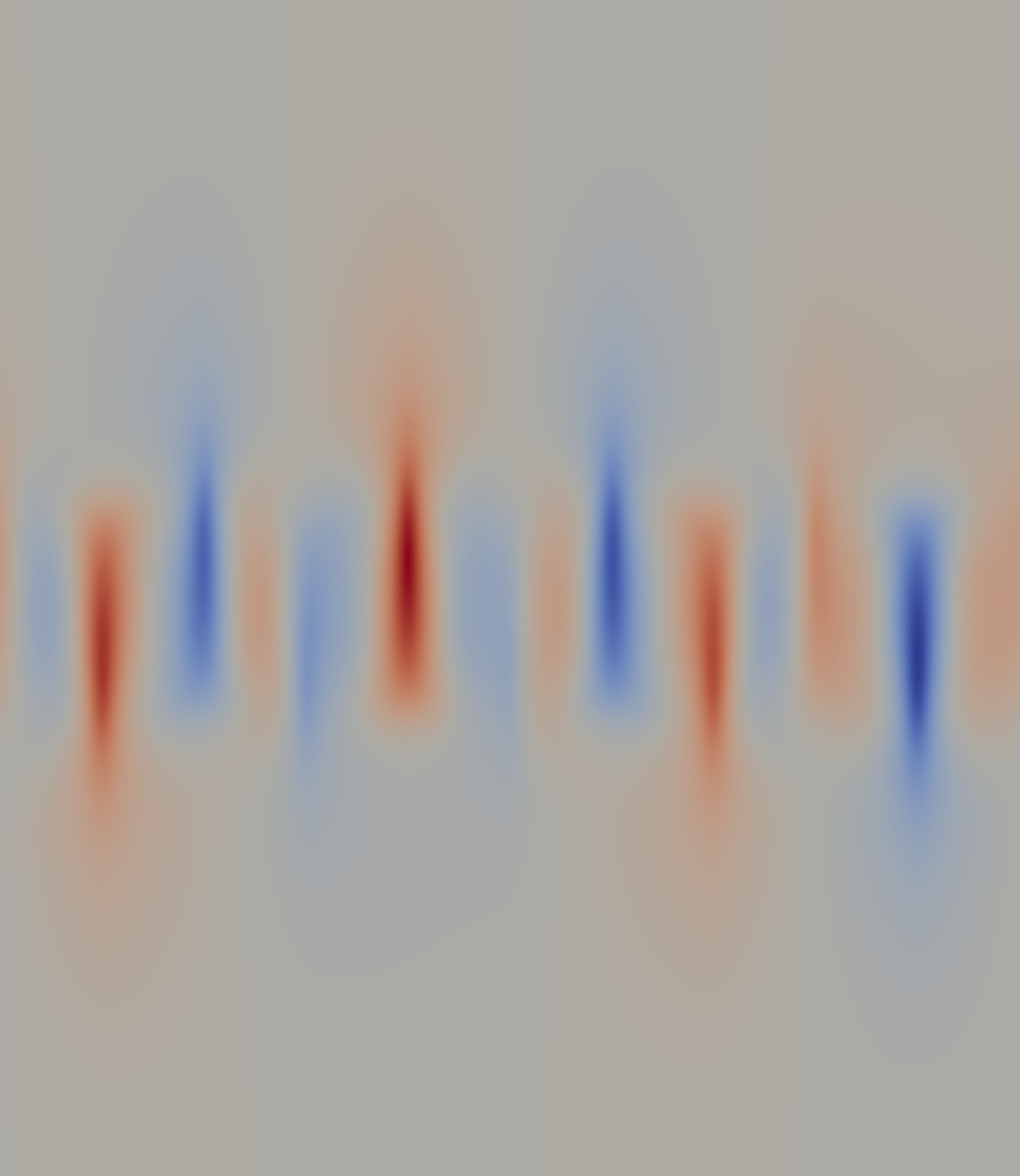} & \includegraphics[width=5.0cm]{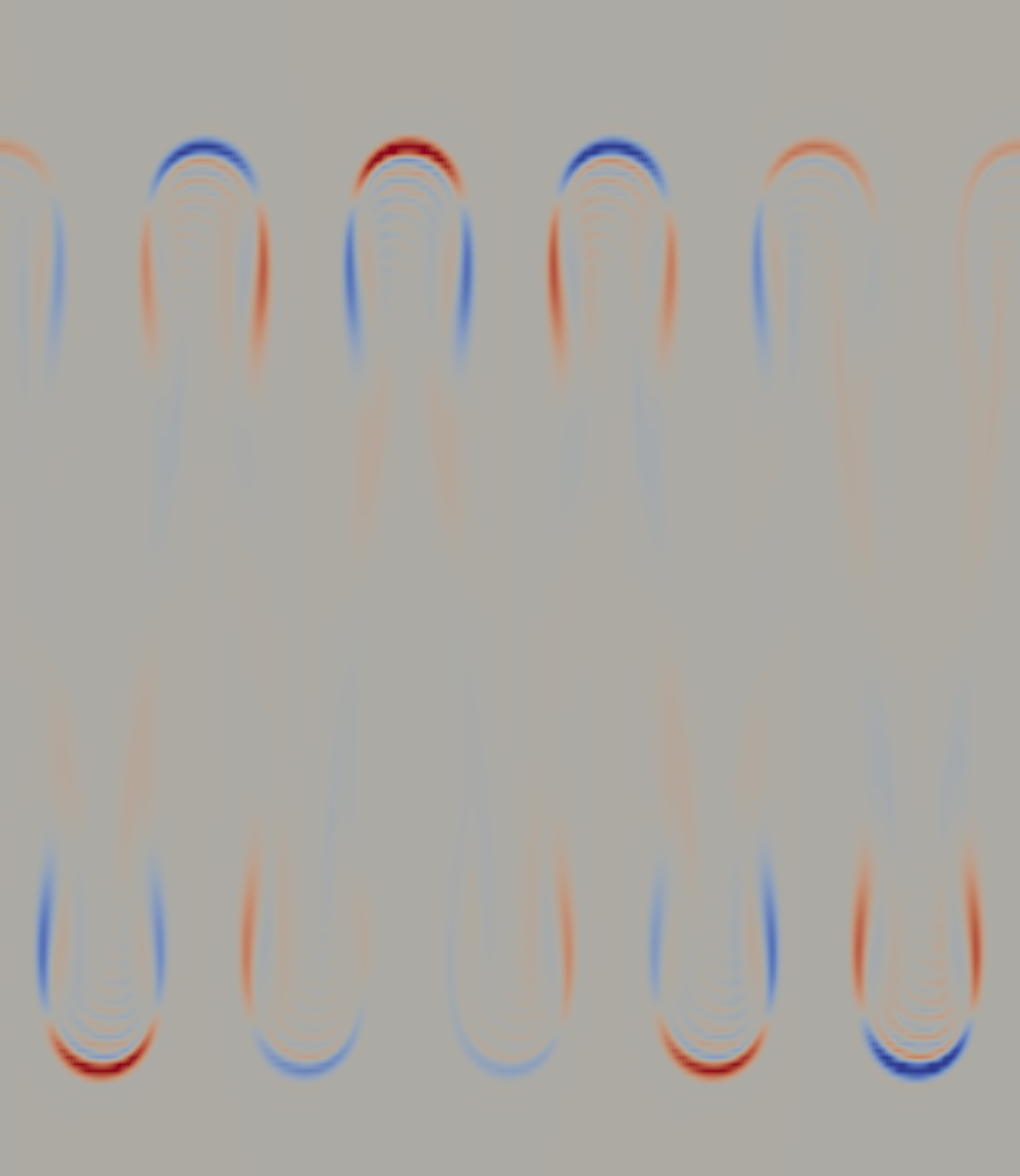}
    \\
    initial salinity perturbation & final salinity perturbation
  \end{tabular}
  \caption{The leading perturbation to the salt fingering system. When the perturbation on the left is applied to the initial
    condition for salinity in the discretised model, the perturbation grows with a growth factor $\sigma \approx 2235$, resulting in a much larger perturbation
  to the final salinity.}
  \label{fig:salt_fingering_ptb}
\end{figure}
\begin{table}[t]
\centering
\begin{tabular}{ccc}
\toprule
       & Runtime (s) & Ratio \\
\midrule
Forward model & 165.89 &     \\
Tangent linear model (averaged) & 65.25  & 1.39 \\
Adjoint model (averaged) & 68.71 & 1.41 \\
\bottomrule
\end{tabular}
\caption{Timings for the salt fingering simulation for computing the perturbation that grows optimally to $T = 0.05$. 
  The optimal perturbation is obtained after $24$ tangent linear and adjoint model solves. 
 The table shows the run time for the forward and the averaged timings for the tangent linear and adjoint solves.
  As can be seen, the tangent linear and the adjoint models take approximately 40\% of the cost of the forward model. The
  optimal ratio is approximately 1.33.}
\label{tab:salt_fingering_timings}
\end{table}

The leading input perturbation is plotted in figure \ref{fig:salt_fingering_ptb}, along with the
resulting linear perturbation to the final state. As visible in the figure, the leading perturbation
encourages the growth of some fingers, while retarding the growth of others. We identified a number
of unstable modes which result in an uneven distribution of salt finger lengths; the physical
mechanism is that longer fingers retard the growth of the shorter fingers since incompressibility
requires a return flow in the opposite direction either side of each finger.  All ten perturbations
computed were found to grow over the time interval $[0, 0.05]$; the leading perturbation grew in
norm by a factor of approximately 2235 over the time window. This secondary instability was first
observed in \cite{mactavish2013}, where these perturbations were activated by the use of unstructured
meshes.

The performance was benchmarked by recording the run times of the forward, tangent linear and
adjoint models on a coarser configuration with a structured mesh of $50 \times 50$ cells and a timestep of
$1 \times 10^{-3}$.  The numerical results can be seen in table~\ref{tab:salt_fingering_timings}.
During the forward solve, the Newton solver typically converges after three iterations.  As both the
adjoint and the tangent linear models replace each Newton solve with one linear solve, a coarse
estimate of the optimal performance is that the tangent linear and adjoint models should take $33\%$
of the run time of the forward model, for an optimal ratio of 1.33. (Efficiency results for derived
models always include the cost of the forward model also, as running the forward model is necessary
to run derived models \citep{naumann2011}).  The numerical results yield a value of approximately $40\%$ of
the cost of the forward model; the tangent linear and the adjoint models approach
optimal performance.

\subsection{Cahn-Hilliard: phase separation}
The Cahn-Hilliard equation is a partial differential equation which describes the process of phase separation, in which
two components of a mixed binary fluid separate to form pure regions of each component \cite{cahn1958}. The equation
has also found applications in image processing, for evolving object contours \cite{capuzzo2002}, and astrophysics,
for modelling the evolution of Saturn's rings \cite{tremaine2003}. The Cahn-Hilliard equation is a nonlinear fourth-order
parabolic equation:
\begin{align}
\frac{\partial c}{\partial t} - \nabla \cdot M \left( \nabla \left( \frac{\mathrm{d}f}{\mathrm{d}c}
- \lambda \nabla^2 c\right) \right) =& \ 0 \quad \mathrm{on}\ \Omega, \\
M \left( \nabla \left( \frac{\mathrm{d}f}{\mathrm{d}c} - \lambda \nabla^2 c\right) \right) =& \ 0 \quad \mathrm{on}\ \partial \Omega, \\
M \lambda \nabla c \cdot n =& \ 0 \quad \mathrm{on}\ \partial \Omega,
\end{align}
where $c$ is the prognostic concentration field ($c=1$ is one fluid, $c=0$ the other), $f$ is the (prescribed) chemical potential, $n$ is the outward unit normal, and
$\lambda$ and $M$ are scalar constants. In order to apply standard continuous finite elements, the fourth-order equation is broken up
into two coupled second-order equations, and a mixed P1-P1 finite element discretisation applied \cite{wells2006}.

\begin{figure}
  \centering
  \begin{tabular}{cc}
    \includegraphics[width=5.0cm]{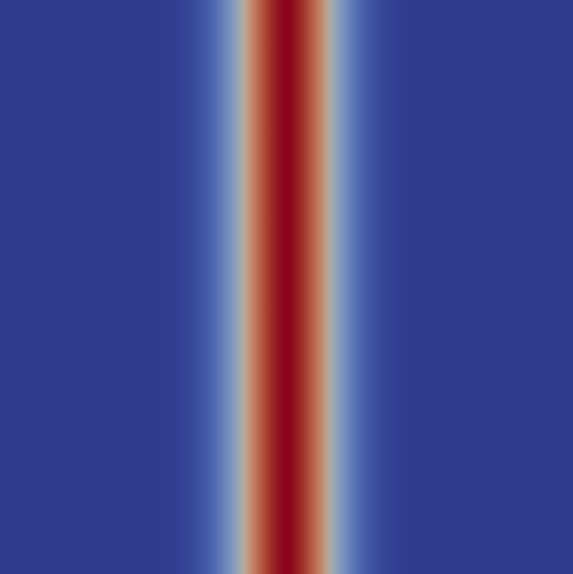} & \includegraphics[width=5.0cm]{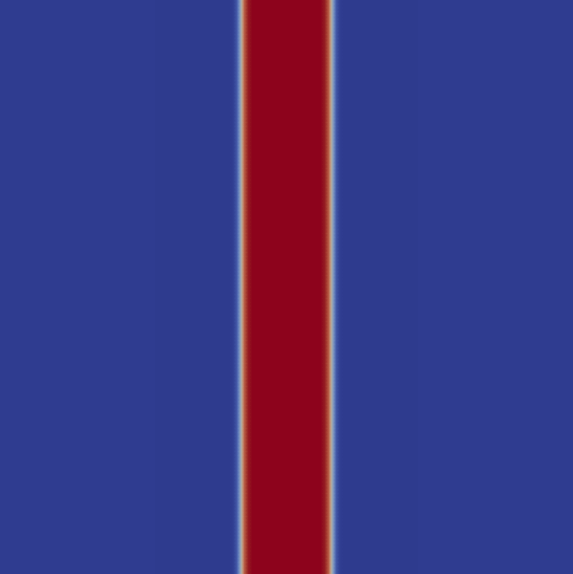}
    \\
    initial concentration & final concentration
  \end{tabular}
  \caption{The initial and final conditions for the Cahn-Hilliard simulation. The color bar ranges from 0 to 1.}
  \label{fig:ch_conditions}
\end{figure}

Generalised stability analysis was employed to investigate the stability of the evolution of the Cahn-Hilliard system from a
randomly perturbed initial condition on the domain $\Omega = \left[0, 2\right]^2$. The initial condition was given by the one-dimensional profile
\begin{equation}
c_0 = c(t=0) = e^{-30 (x - 1)^2}.
\end{equation}
The constants were set to $\lambda = 10^{-2}$ and $M = 1$, and $f = 100c^2(1-c)^2$. The initial (at $t=0$) and final
conditions (at $t=5\times10^{-4}$) for the simulation are presented in figure \ref{fig:ch_conditions}. The mesh had 150 elements in both
the $x-$ and $y-$ directions, leading to a mixed function space with 90602 degrees of freedom. The timestep $\Delta t$ was set to
$5\times10^{-6}$. The simulations were run in parallel across 8 cores using MPI.

\begin{figure}
  \centering
  \begin{tabular}{c}
    \includegraphics[width=10.0cm]{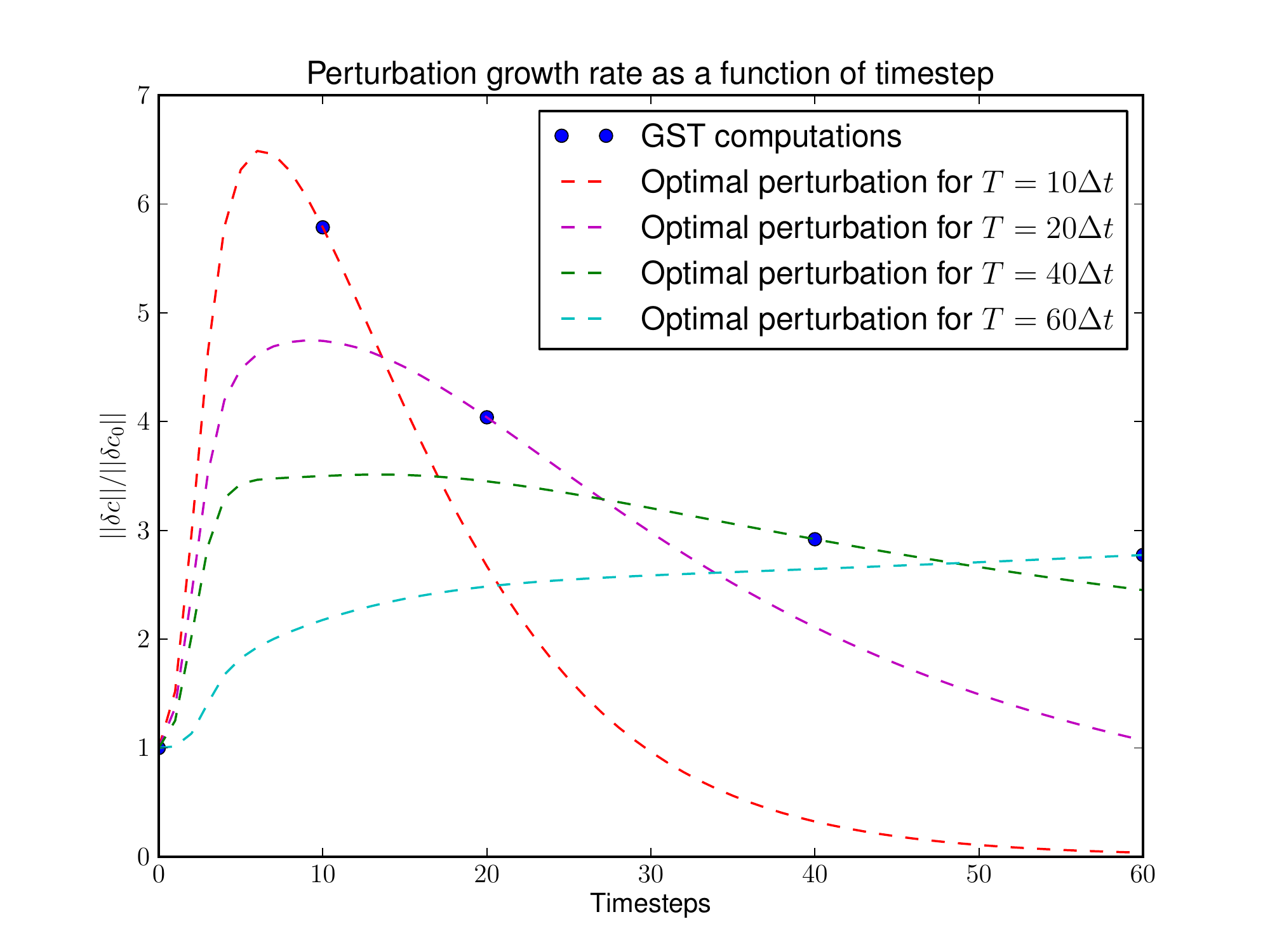}
  \end{tabular}
  \caption{The growth rate of the optimal perturbation computed using GST at various times (blue
  dots), and the growth rate of the optimal perturbation associated with various timesteps, computed
using the nonlinear model (dashed lines). Note that the choice of $T$ is crucial. To compute the dashed curves, the identified perturbation
was scaled to have norm $\left|\left| \delta c_0 \right| \right| = 10^{-7}$, and was added to unperturbed initial 
condition. The nonlinear model was then executed with this perturbed initial condition, and the
results compared to the original unperturbed nonlinear trajectory. The fact that the dashed curves
(observed from the nonlinear model) match the GST predictions indicates that the GST analysis is
correct.}
  \label{fig:ch_growths}
\end{figure}

\begin{figure}
  \centering
  \begin{tabular}{cc}
    \includegraphics[width=5.0cm]{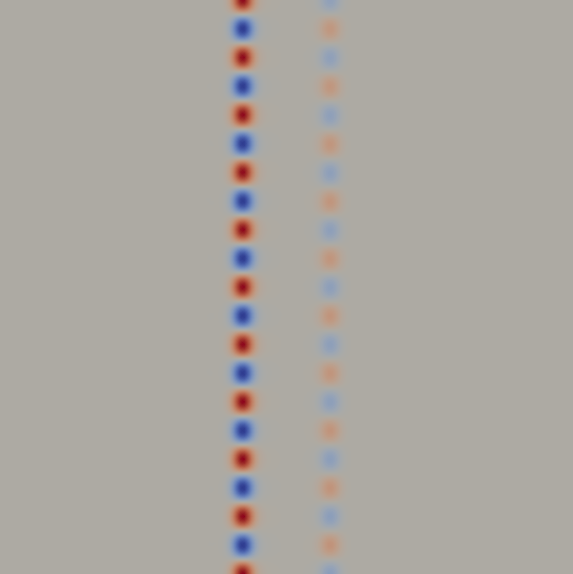} & \includegraphics[width=5.0cm]{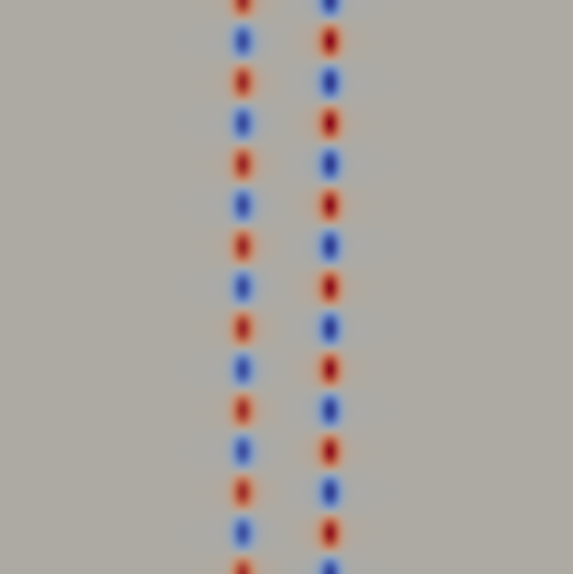}
    \\
    $T = 10\Delta t$ & $T = 20 \Delta t$
    \\
    \includegraphics[width=5.0cm]{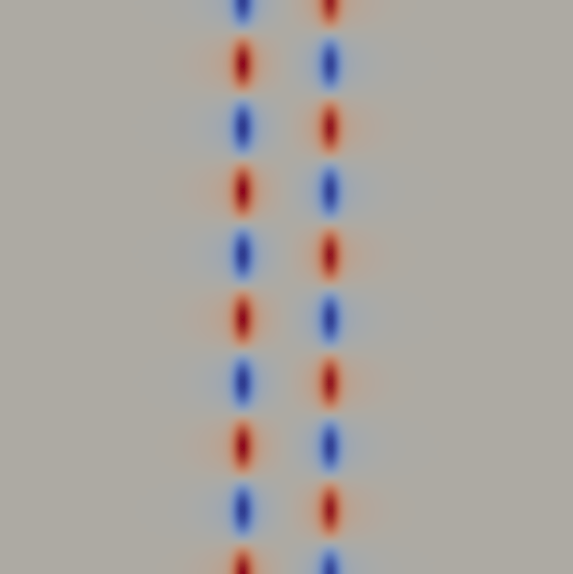} & \includegraphics[width=5.0cm]{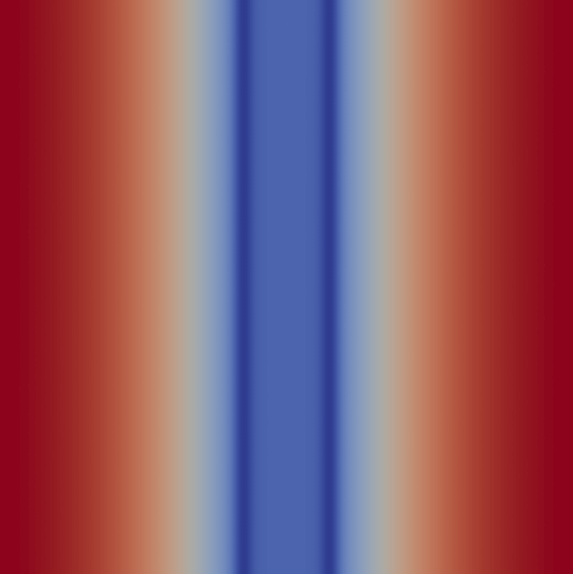}
    \\
    $T = 40\Delta t$ & $T = 60 \Delta t$
  \end{tabular}
  \caption{The perturbation to the Cahn-Hilliard concentration that grows optimally (equivalently, the leading singular vector of
  the propagator), displayed for various integration periods. As the propagator is linear by definition, the scales of the perturbations
  do not matter, and so the perturbations are normalised to have unit norm. The optimal perturbation clearly depends on the
  time for which the propagator is defined.}
  \label{fig:ch_gst}
\end{figure}
\begin{table}[t]
\centering
\begin{tabular}{ccc}
\toprule
       & Runtime (s) & Ratio \\
\midrule
Forward model & 66.63  &     \\
Tangent linear model (averaged) & 17.64 & 1.26 \\
Adjoint model (averaged) & 17.92  & 1.27 \\
\bottomrule
\end{tabular}
\caption{Timings for the Cahn-Hilliard simulation for computing the perturbation that grows optimally to $T = 10\Delta t$. 
  The perturbation is obtained after $72$ tangent linear and adjoint model solves. 
 The table shows the run time for the forward and the averaged timings for the tangent linear and adjoint solves. The optimal ratio is
 approximately 1.25.}
\label{tab:cahn-hilliard-timings}
\end{table}

The generalised stability analysis was used to compute the optimally linearly growing perturbations to the initial condition for concentration and their
growth rates at times
$T = 10\Delta t,$ $20 \Delta t,$ $40 \Delta t,$ and $60 \Delta t$. The optimal growth rates computed using GST for these values of $T$ are shown in figure \ref{fig:ch_growths}
(solid blue dots).
In general, the perturbation that grows optimally to a time $T_1$ will be different to the perturbation that grows optimally to a time $T_2 \ne T_1$; that is,
the singular vectors are sensitive to the integration period of the propagator \cite[pg. 220]{kalnay2002}. This is indeed the case for the GST analysis of the
Cahn-Hilliard system. The leading singular vectors of the propagator defined with respect to various times is shown in figure \ref{fig:ch_gst}. 

To further verify the utility of GST, the nonlinear model was perturbed with each identified optimal perturbation, in order to compare the growth rates
predicted by the GST with the actual growth rates observed. The predictions and observations match closely, indicating that the
GST is indeed predicting the quantitative behaviour of the system (figure \ref{fig:ch_growths}, dashed lines). The growth curves of the
perturbations demonstrate the phenomenon of transient growth: initial growth in magnitude over some finite time horizon, followed by
asymptotic decay. Such phenomena are characteristic of nonnormal systems \cite{trefethen2006}.

The run times of the forward, tangent linear and adjoint models for the setup with $T=10\Delta t$ are shown in table~\ref{tab:cahn-hilliard-timings}. 
For this configuration, the Newton solver typically converges after four iterations during the forward simulation. 
Therefore, the optimal performance can be estimated to be $25\%$ of the run time of the forward model, for an optimal ratio of 1.25.
The benchmark results yield a value of $27\%$ of the cost of the forward model; the tangent linear and adjoint models approach optimal efficiency.

\subsection{Gross-Pitaevskii: soliton solutions}

The Gross-Pitaevskii equation \cite{gross1961,pitaevskii1961} is a nonlinear Schr\"odinger equation that describes the dynamics
of a quantum system of identical bosons. The nondimensional equation governing the evolution of the wavefunction $\Psi$ is given by
\begin{equation}
  i\frac{\partial \Psi}{\partial t} + \nabla^2\Psi + s|\Psi|^2\Psi = 0,
\end{equation}
where $s$ is a parameter ($s = 1$ is the focussing case, $s = -1$ the defocussing case). In particular, the Gross-Pitaevskii equation
describes the behaviour of Bose-Einstein condensates, a state of matter observed when a dilute gas of bosons is cooled to temperatures
close to absolute zero \cite{bose1924,einstein1924}. Bose-Einstein condensates are of considerable interest as they permit
black hole analogues: systems from which acoustic perturbations, rather than light, are unable to escape \cite{unruh1981}. This
could potentially allow the laboratory-scale experimental investigation of the physics of black holes \cite{farrell2008,lahav2010}.

Generalised stability theory was employed to investigate the stability of the one-dimensional soliton solution of the focussing Gross-Pitaevskii
equation
\begin{equation} \label{eqn:soliton}
\Psi = \sqrt{2} \frac{\exp{(\frac{i}{2} x + \frac{3i}{4}t)} }{\cosh{(x - t)}}
\end{equation}
to perturbations in the initial condition. The Gross-Pitaevskii equation was solved with piecewise linear finite elements on the domain $\Omega = [-10, 10]$
with periodic boundary conditions applied. The initial condition was achieved by pointwise evaluation of \eqref{eqn:soliton}, and the equations were
advanced in time from $0$ to $T$ using the implicit midpoint rule. The interval was discretised with
$N = 480$ elements, and the timestep was set to $\Delta t = 0.03125$.

\begin{figure}
  \centering
  \begin{tabular}{cc}
    \includegraphics[width=6.5cm]{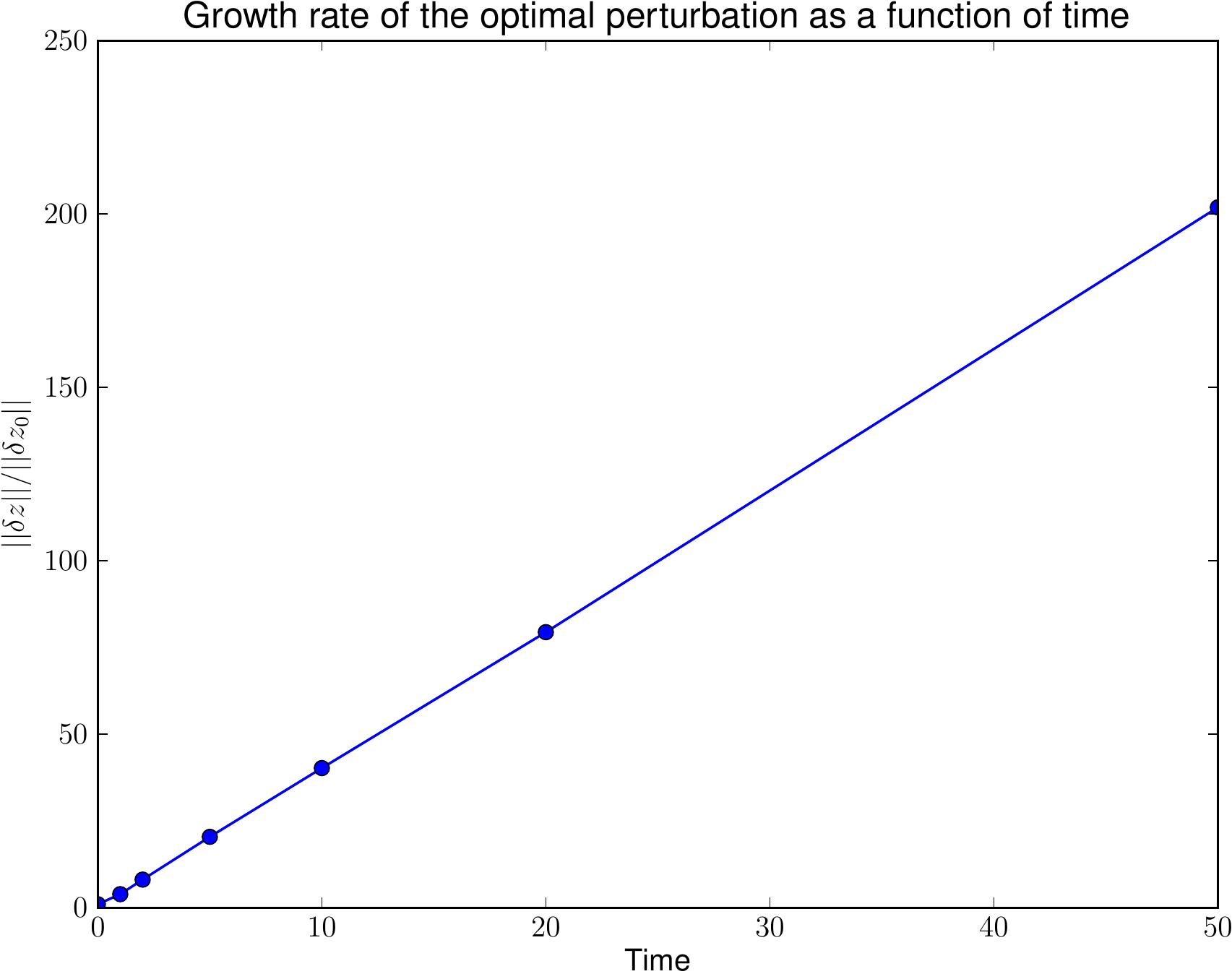} &
    \includegraphics[width=7.15cm]{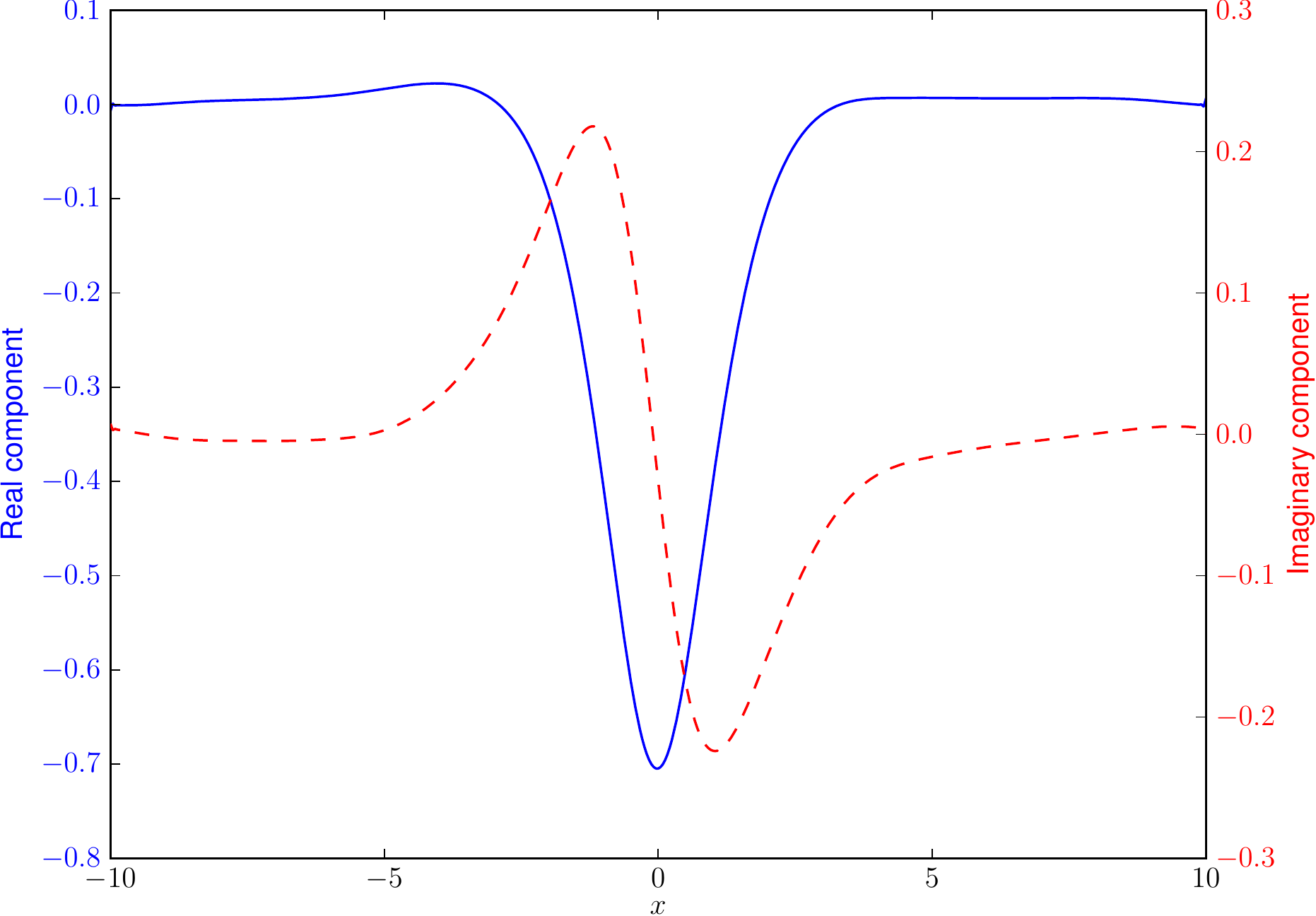} \\
    (a) & (b)
  \end{tabular}
  \caption{(a): The growth rate of the optimal perturbation to Gross-Pitaevskii system as a function of time. The optimal perturbations associated with times $T > 10$ are very similar. The linear growth of this perturbation was verified using the original nonlinear model up to $T=500$. (b): The optimal perturbation associated with time $T = 50 \Delta t$. The solid blue line is the real component, while the dashed red line is the imaginary component.}
  \label{fig:gp_growths}
\end{figure}

The results of the GST calculation for various times are shown in
figure \ref{fig:gp_growths}a. In this example, approximately linear
growth of the optimal perturbations is observed. For $T>10$, all GST
calculations yielded very similar perturbations (figure \ref{fig:gp_growths}b shows the
perturbation for T=$50\Delta t$). 

\begin{figure}
  \centering
  \begin{tabular}{c}
    \includegraphics[width=10.0cm]{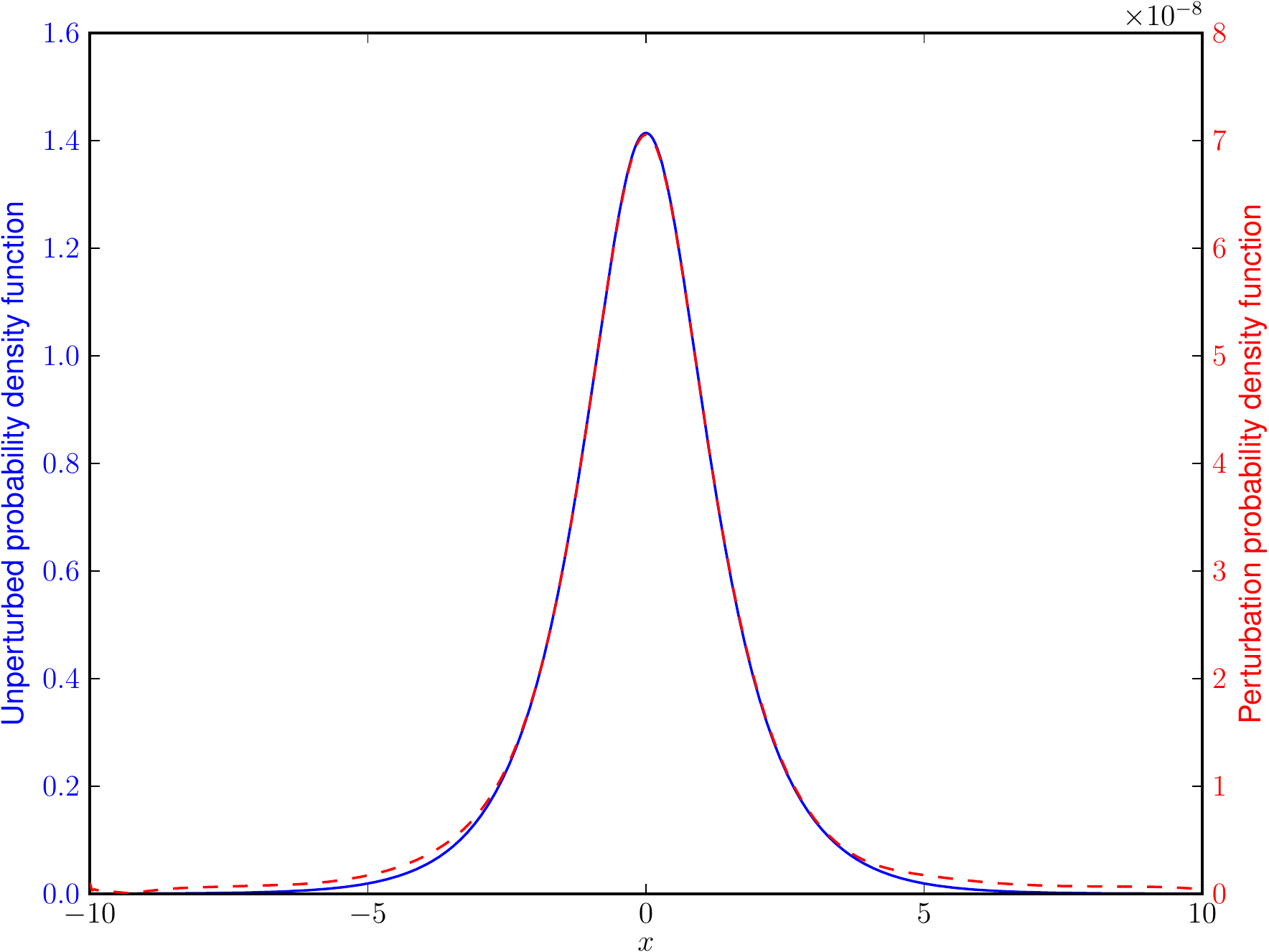}
  \end{tabular}
  \caption{The probability density functional for the unperturbed
    Gross-Pitaevskii soliton initial condition (solid blue line) and
    the optimal perturbation associated with times $T > 10$ (red
    dashed line). The perturbation corresponds to shifting to a higher
    (or lower, with negative coefficient) amplitude soliton solution;
    this is evident since the perturbation has almost the same shape
    as the soliton itself, but with slightly wider support. Higher
    (lower) amplitude soliton solutions have greater (lesser) speeds,
    and so the growth rate is linear in time.}
  \label{fig:gp_pdfs}
\end{figure}

This optimal perturbation corresponds to
shifting along the family of soliton solutions parameterised by their amplitude. Since each
member of this family has a different speed, perturbing in this
direction leads to a similar shaped soliton moving at a different
speed, hence the linear growth in the perturbation. This indicates that the soliton
solutions are stable. This is illustrated in figure \ref{fig:gp_pdfs}.

The timing results are given in table~\ref{tab:gross-pitaevskii-timings}. 
For this example, the model only has one spatial dimension which makes the linear solves computationally cheap.
As a consequence, the cost of the linear solves does not dominate the cost of the symbolic manipulation for low resolutions ($N = 480$),
and so the efficiency ratio is suboptimal.
However, as the mesh resolution is increased ($N = 12,000$), the cost of the linear solves increases while the cost of the symbolic manipulation 
does not. Therefore as the mesh is refined, the efficiency ratio approaches the optimal value. Of course, for one-dimensional problems, such fine
discretisations are often unnecessary; however, the asymptotic regime is rapidly reached for problems of two or more dimensions, as in the previous
examples.

\begin{table}[t]
\centering
\begin{tabular}{c|cc|cc|cc}
\toprule
Mesh elements & \multicolumn{2}{|c|}{$N = 480$} & \multicolumn{2}{|c}{$N = 6,000$} & \multicolumn{2}{|c}{$N = 12,000$} \\ 
              & Runtime (s) & Ratio  & Runtime (s) & Ratio & Runtime (s) & Ratio \\ 
\midrule
Forward model & 11.84 &  & 58.06 & & 109.67  \\ 
TLM (averaged) & 23.88 & 3.02 & 47.13 & 1.81 & 55.44 & 1.51 \\ 
      ADM (averaged) & 24.50 & 3.07 & 51.63 & 1.89 & 58.88 & 1.54 \\ 
\bottomrule
\end{tabular}
\caption{Timings for the Gross-Pitaevskii simulation for computing the perturbation that grows optimally to $T = 10$. 
  The perturbation is obtained after $16$ tangent linear (TLM) and adjoint model (ADM) solves. 
 The table shows the run time for the forward and the averaged timings for the tangent linear and adjoint solves.
 The Newton solver converges on average after two Newton iterations, which means that the optimal ratio is approximately 1.5.
 With low resolution ($N = 480$), the cost of the linear solves does not dominate the symbolic
 manipulation; as the mesh is refined ($N = 12,000$), the linear solves become the
 dominant cost, and the efficiency ratio approaches the optimal value.
} 
\label{tab:gross-pitaevskii-timings}
\end{table}

\section{Conclusions}
Generalised stability theory is a powerful tool for investigating the dynamics of physical systems, but the
difficulty of implementing it has been a major impediment to its widespread application. The core contribution
of this paper has been to remove this barrier. By employing a new high-level symbolic approach to automating
the derivation of adjoint and tangent linear models, conducting a generalised stability analysis is now
straightforward, even for parallel discretisations of complex nonlinear coupled time-dependent problems. The widespread
applicability of the framework was demonstrated on examples drawn from geophysical fluid dynamics, phase
separation, and quantum mechanics.

Adjoint and tangent linear models arise across computational mathematics, not merely in stability
analysis. Therefore, the same core technology of the automated derivation of adjoint and tangent
linear models has major applications in optimisation constrained by partial differential equations,
automated error analysis and goal-based adaptivity, continuation and bifurcation analysis, data
assimilation, and uncertainty quantification.

A further setting where adjoints prove very useful is Markov Chain Monte Carlo (MCMC) algorithms that
are used for Bayesian inference problems.  It has been shown that if the derivative of the
observation model is available, then the convergence of the algorithm is considerably faster
\cite{roberts1996,martin2012}. The derivative is also useful for avoiding getting stuck in local maxima
\cite{beskos2011}. Bayesian inverse problems have been recently rigorously formulated on function
spaces in a well-posed manner; this means that MCMC algorithms can be appropriately modified so that
the number of iterations required to converge is independent of mesh resolution \cite{cotter2010}.
The possibility of automated adjoint generation opens up the possibility of applying these
algorithms in a very broad range of applications where they would not otherwise reach.

Another area of particular relevance to this work is the application of techniques from optimal
control to transient growth and bypass transition: whereas generalised stability theory accounts
for nonnormal effects, such analyses account for both nonnormal and nonlinear effects
\cite{monokrousos2011,juniper2011b,juniper2011a}.  These techniques rely fundamentally on the
solution of the associated adjoint system to provide the gradient information necessary for the
nonlinear optimisation.  Future work will be to explore these applications, and extend these
techniques to physical systems where their implementation was previously impractical.

\section*{Acknowledgements}
PEF's work was supported by EPSRC grants EP/I00405X/1 and EP/K030930/1 and a Center of Excellence grant from the
Norwegian Research Council to the Center for Biomedical Computing at Simula Research Laboratory.
SWF's work was supported by EP/I00405X/1, the Grantham Institute for Climate Change and Fujitsu Laboratories of Europe Ltd.
The secondary instability in salt fingers was originally discovered by F. P. MacTavish.  The authors
would like to thank G. N. Wells for providing the Cahn-Hilliard solver, and the anonymous reviewers for
their thorough and constructive reviews. 

\bibliographystyle{siam}
\bibliography{literature}
\end{document}